\documentclass[12pt]{article}
\usepackage{cite}
\renewcommand{\theequation}
{\arabic{section}.\arabic{equation}}

\makeatletter
\def\eqnarray{ \stepcounter{equation} \let\@currentlabel=\theequation
 \global\@eqnswtrue
 \global\@eqcnt\z@
 \tabskip\@centering
 \let\\=\@eqncr
 $$\halign to \displaywidth\bgroup\@eqnsel\hskip\@centering
 $\displaystyle\tabskip\z@{##}$&\global\@eqcnt\@ne
 \hfil$\displaystyle{{}##{}}$\hfil
 &\global\@eqcnt\tw@$\displaystyle\tabskip\z@{##}$\hfil
 \tabskip\@centering&\llap{##}\tabskip\z@\cr}
\makeatother

\makeatletter
\def\@arrayacol{\edef\@preamble{\@preamble \hskip .2\arraycolsep}}
\def\array{\let\@acol\@arrayacol \let\@classz\@arrayclassz
\let\@classiv\@arrayclassiv \let\\\@arraycr\def\@halignto{}\@tabarray}
\makeatother
\renewcommand{\arraystretch}{1.6}
\makeatletter                                                              
\newcounter{subeqncnt} 
\def\thesubeqncnt{\alph{subeqncnt}}
\def\subequations{\begingroup% 
   \stepcounter{equation}\edef\@tempa{\theequation}%
   \let\c@equation\c@subeqncnt\c@subeqncnt\z@
   \edef\theequation{\@tempa\noexpand\thesubeqncnt}}

\makeatother
\newcommand{\del}{\partial}

\newcommand{\rf}[1]{(\ref{#1})}
\newcommand{\bea}{\begin{eqnarray}}
\newcommand{\eea}{\end{eqnarray}}

\newcommand{\ep}{\varepsilon}

\newcommand{\dd}{{\rm d}}

\newcommand{\prt}{\partial}
\newcommand{\half}{\frac{1}{2}}

\newcommand{\auxA}{{\phi}}

\newcommand{\hd}{\bar{d}}%\hat{d}}
\newcommand{\hZ}{\hat{Z}}%Z

\newcommand{\UvUa}{${\rm U}(1)_{\rm V} \!\times\! {\rm U}(1)_{\rm A}$}

\newcommand{\tinyspace}{\hspace{0.6pt}}
\begin{document}

\setlength{\baselineskip}{7mm}
\begin{titlepage}
\begin{flushright}
{\tt DIAS-STP-03-07} \\
{\tt TIT/HEP-508} \\
August, 2003
\end{flushright}
 
\vspace{2cm}

\begin{center} 
{\Large Quantization of Bosonic String Model \\
in 26+2-dimensional Spacetime} 

\vspace{1cm}

{\sc{Takuya Tsukioka}}$^*$
and 
{\sc{Yoshiyuki Watabiki}}$^\dagger$\\
$*${\it{School of Theoretical Physics,}} 
{\it{Dublin Institute for Advanced Studies,}} \\
{\it{10 Burlington Road, Dublin 4, Ireland}} \\
{\sf{tsukioka@synge.stp.dias.ie}} \\
and \\
$\dagger${\it{Department of Physics,}} 
{\it{Tokyo Institute of Technology,}} \\
{\it{Oh-okayama, Meguro, Tokyo 152, Japan}} \\
{\sf{watabiki@th.phys.titech.ac.jp}}
\end{center}

\vspace{1cm}

\begin{abstract}
We investigate the quantization of the bosonic string model 
which has a local {\UvUa} gauge invariance  
as well as the general coordinate 
and Weyl invariance on the world-sheet. 
The model is quantized by Lagrangian and Hamiltonian BRST formulations 
{\it \'a la} Batalin, Fradkin and Vilkovisky 
and noncovariant light-cone gauge formulation. 
Upon the quantization the model turns out to be formulated 
consistently in 26+2-dimensional background spacetime 
involving two time-like coordinates.  
\end{abstract}

\end{titlepage}

%%%%%
\section{Introduction}
\setcounter{equation}{0}
\setcounter{footnote}{0}

It is the purpose of this paper to cast some further light 
upon constructions of theories involving two time coordinates. 
To consider the physics which has more than two time coordinates 
might be a clue to understand the origin of time and spacetime. 

Several theories constructed on spacetime with two time coordinates 
are investigated from various viewpoints, 
such as F-theory~\cite{v}, two-time physics~\cite{b} 
and 12-dimensional super Yang-Mills theory~\cite{ns}. 
F-theory is proposed by Vafa 
as an extended concept of string theory 
and constructed by using field theory of super (2,2)-brane~\cite{hp} 
with 10+2-dimensional background spacetime. 
The two-time physics is proposed by Bars as a device 
for searching a unified theory 
and developed by himself and his collaborators~\cite{two-time}. 
In this context, string-particle systems are 
proposed~\cite{bdkm} from string theory point of view. 
By introducing constant null vectors in background spacetime 
into the formulation, 
the 12-dimensional super Yang-Mills theory~\cite{ns}  
is also proposed. 

Some years ago, one of the authors (Y.W.) had proposed a model 
which has a {\UvUa} gauge symmetry in two-dimensional 
spacetime~\cite{wata1}. 
The striking feature of this model is that 
there exists a negative norm state in two-dimensional spacetime  
as the same as string theories~\cite{wata1}. 
Using the {\UvUa} gauge symmetry he also proposed string models 
which have two time-like coordinates in ref.~\cite{wata2}. 
These models have the {\UvUa} gauge symmetry or a supersymmetric 
version of the {\UvUa} gauge symmetry on the two-dimensional world-sheet. 
The background spacetimes of the {\UvUa} bosonic and superstring model 
might be 26+2 and 10+2 dimensions, respectively. 
In ref.~\cite{wata3} manifest covariant formulations of 
the string models are given. 

We in this paper further study the {\UvUa} string model. 
In particular, it would be obviously important to 
explicitly carry out the quantization, 
so that we can argue not only the critical dimension 
but also the mass spectrum at the quantum level. 
Since many concepts in string theories are presented in bosonic models, 
we focus our attentions on the bosonic {\UvUa} string model in this paper. 
A quantization of the superstring model based on our framework 
will be discussed in an additional work elsewhere~\cite{tw}. 

The {\UvUa} string model is constructed as gauge field theory 
on two-dimensional world-sheet~\cite{wata2}. 
Although the similar models were investigated in refs.~\cite{bdkm,dg}, 
an advantage of the formulation of our model is 
its manifest covariant expression in the background spacetime 
by using the {\UvUa} gauge symmetry~\cite{wata3}, 
so that in this paper we can easily carry out the quantization 
with preserving the covariance.  
The {\UvUa} gauge symmetry is essential in our model. 
In constructing the covariant action, 
the generalized Chern-Simons action~\cite{kawata} 
proposed by Kawamoto and one of the authors (Y.W.) 
as a new type of topological action 
plays an important key role. 

There are two remarks in quantizing our model. 
Firstly the action has a reducible symmetry 
which originally arises from symmetric structures 
of the generalized Chern-Simons action~\cite{kostu}. 
Secondly the gauge algebra is open. 
In the covariant BRST quantization of the system 
including reducible and open gauge symmetry, 
we need to use the formulations developed by 
Batalin, Fradkin and Vilkovisky~\cite{bv,bfv}. 
By adopting these methods 
we explicitly show the covariant quantizations are successfully 
carried out in the Lagrangian and the Hamiltonian formulations. 

In order to treat the dynamics of our model more directly, 
we also quantize the model in noncovariant light-cone gauges. 
The suitable noncovariant gauge conditions can be imposed by 
residual symmetries of the {\UvUa} gauge symmetry 
and we can then solve all of the gauge constraints explicitly. 
We can also confirm that the existence of two time-like coordinates 
is not in conflict with the unitarity of the theory, 
since the two time-like coordinates are required 
by our ``gauge'' symmetry. 

As an important feature of quantum string models,  
we can argue the critical dimension of the background spacetime. 
In usual bosonic string theories, the critical dimension is 
25+1~\cite{gsw,bh,p}, which is estimated by the BRST~\cite{ko,h} 
and the light-cone gauge formulation~\cite{ggrt}. 
For our bosonic model, the critical dimension turns out to be 26+2. 
We obtain this result directly from both the BRST and the noncovariant 
light-cone gauge formulations. 

This paper is organized as follows: 
We first introduce the {\UvUa} string model and explain semiclassical 
aspects of the model in Section 2. 
The preparation for the quantization is also given in this section.   
We present the covariant quantization based 
on the Lagrangian formulation in Section 3. 
In this section we investigate the perturbative aspect of the 
quantized model and determine the critical dimension of 
our {\UvUa} string model. 
In Section 4 the covariant quantization of the same model is 
carried out in the Hamiltonian formulation.  
By taking suitable gauge fixing conditions 
we reproduce the same gauge-fixed action in the Lagrangian formulation. 
We also obtain the BRST charge in this section.  
The quantization under noncovariant light-cone gauge 
fixing conditions is carried out in Section 5. 
We then study the symmetry of the background spacetime and 
obtain the same critical dimension by direct computation 
of the full quantum Poincar\'e algebra. 
We also present a mass-shell relation of the model and 
give low energy quantum states.  
Conclusions and discussions are given in the final section. 
Appendixes A and B contain our conventions. 
We also exhibit the BRST formulation of {\UvUa} model 
without two-dimensional gravity in Appendix C.  

%%%%%
\section{{\UvUa} bosonic string model}
\setcounter{equation}{0}
\setcounter{footnote}{0}

\subsection{Classical action and its symmetries}

The {\UvUa} bosonic string model~\cite{wata2} 
described by two-dimensional field theory 
consists of 
$D$ scalar fields $\xi^I(x)$, 
an Abelian gauge field $A_m(x)$ and the metric $g_{mn}(x)$. 
The two-dimensional spacetime coordinates are 
$x^m$ ($m \!=\! 0$, $1$) and 
the signature of metric is $(-,+)$. 
Our conventions are given in Appendix A.  
The scalar fields $\xi^I(x)$ are considered to be 
string coordinates in $D$-dimensional flat background spacetime 
with the background metric: 
\begin{equation}
\label{FlatmetricDefinitionGravity}
\eta_{IJ}  \ = \  \eta^{IJ}  \ = \ 
   \left\{\begin{array}{cl}
       -1   &  \hspace{36pt}\hbox{($I=J=0$)}           \\
        1   &  \hspace{36pt}\hbox{($I=J=i$, \ $i=1$, $2$, \ldots, $D-3$)} \\
       -1   &  \hspace{36pt}\hbox{($I=J=\widehat{0}$)} \\
        1   &  \hspace{36pt}\hbox{($I=J=\widehat{1}$)} \\
        0   &  \hspace{36pt}\hbox{(otherwise)}
   \end{array}\right.
\end{equation}
The indices $I$ and $J$ run through 
$0$, $1$, $2$, \ldots, $D-3$, $\widehat{0}$, $\widehat{1}$. 
As we will explain, the unitarity as a two-dimensional field theory 
requires two negative signatures to 
the background metric $\eta_{IJ}$, 
because the ${\rm U}(1)_{\rm A}$ gauge transformation 
as well as the general coordinate transformations
removes a negative norm state. 
At the quantum level 
the absence of conformal anomaly requires $D \!=\! 28$, 
however, we need not specify the value of $D$ at the classical level. 

The covariant action of the present model~\cite{wata3} is  
\begin{equation}\label{LagrangianBosonicGravity}
S = \int\! \dd^2 x \, \sqrt{-g} \, \bigg(\!
- \half \, g^{mn} \prt_m \xi^I \prt_n \xi_I 
+ \tilde{A}^m \auxA_I \prt_m \xi^I
\, \bigg) 
 + S_{\rm GCS},   
\end{equation}
where 
$$
\begin{array}[b]{rclcrcl}
g(x) &=& \det g_{mn}(x),  
&\qquad&
\sqrt{-g(x)} \, \tilde{A}^m(x) &=& \ep^{mn} A_n(x). 
\end{array}
$$
The action $S_{\rm GCS}$ is 
the generalized Chern-Simons action which is formulated in 
two-dimensional spacetime~\cite{kawata} 
\begin{equation}\label{LagrangianGCSoriginal}
S_{\rm GCS}  =  \int\! \dd^2 x \, \sqrt{-g} \, \bigg(
\tilde{B}^{mI} \prt_m \auxA_I - \half \tilde{C} \auxA^I \auxA_I 
\bigg), 
\end{equation}
where 
$$
\begin{array}[b]{rclcrcl}
\sqrt{-g(x)} \, \tilde{B}^{mI}(x) &=& \ep^{mn} B^I_n(x) 
\, , &\qquad&
\sqrt{-g(x)} \, \tilde{C}(x) &=& \half \, \ep^{mn} C_{mn}(x) 
\, . 
\end{array}
$$
The fields $\auxA^I(x)$, $B_m^I(x)$ and $C_{mn}(x)$ 
are introduced for the purpose that 
the action $S$ has the covariant form in the background spacetime. 
A derivation of the action (\ref{LagrangianGCSoriginal}) 
from the original generalized Chern-Simons action 
has been given in the paper~\cite{wata3}. 

The action (\ref{LagrangianBosonicGravity}) is invariant 
under the following {\UvUa} gauge transformations, 
\begin{eqnarray}
\delta\xi^I 
&=& v'\phi^I, 
\nonumber \\
\delta\tilde{A}^m 
&=& \frac{\ep^{mn}}{\sqrt{-g}} \del_n v + g^{mn} \del_n v', 
\nonumber\\ 
\delta\tilde{B}^{mI} 
&=& - \, v \frac{\ep^{mn}}{\sqrt{-g}} \del_n \xi^I + v' g^{mn} \del_n \xi^I, 
\label{u1u1} \\
\delta\tilde{C} 
&=& \del_m v' \tilde{A}^m - v' \nabla_m \tilde{A}^m, 
\nonumber \\ 
\delta \phi^I 
&=& \delta g_{mn} = 0, 
\nonumber 
\end{eqnarray}
where the parameters $v(x)$ and $v'(x)$ correspond to the 
vector ${\rm U}(1)$ transformation ``${\rm U}(1)_{\rm V}$'' and the 
axial vector ${\rm U}(1)$ transformation ``${\rm U}(1)_{\rm A}$'', 
respectively. 
Since the generalized Chern-Simons action is invariant under  
nontrivial gauge transformations,  
the action (\ref{LagrangianBosonicGravity}) 
is also invariant under these gauge transformations with 
gauge parameters $u^I(x)$ 
and $\sqrt{-g(x)} \, \tilde{w}^m(x) \equiv \ep^{mn}w_n(x)$,  
\begin{eqnarray}
\delta\tilde{B}^{mI} 
&=& \frac{\ep^{mn}}{\sqrt{-g}} \del_nu^I - \tilde{w}^m \phi^I, 
\nonumber \\
\delta\tilde{C}
&=& \nabla_m\tilde{w}^m, 
\label{gcs_trans}\\
\delta \xi^I 
&=& 
\delta \tilde{A}^m = \delta \phi^I = \delta g_{mn} = 0. 
\nonumber 
\end{eqnarray}
The action (\ref{LagrangianBosonicGravity}) is invariant under 
the general coordinate transformations and the Weyl transformation  
\begin{equation}
\begin{array}{rcl}
\delta\xi^I &=& k^n \del_n \xi^I,  
\\
\delta\tilde{A}^m &=& k^n \del_n \tilde{A}^m 
                      - \del_n k^m \tilde{A}^n 
                      + 2 s \tilde{A}^m,  
\\
\delta\tilde{B}^{mI} &=& k^n \del_n \tilde{B}^{mI}
                         - \del_n k^m \tilde{B}^{nI} 
                         + 2 s \tilde{B}^{mI}, 
\\
\delta\phi^I &=& k^n \del_n \phi^I,  
\\
\delta\tilde{C} &=& k^n \del_n \tilde{C} + 2 s \tilde{C}, 
\\
\delta g_{mn} &=& 
   k^l \del_l g_{mn} + \del_m k^l g_{ln} + \del_n k^l g_{ml}
   - 2 s g_{mn}, 
\label{general_coordinate_Weyl_trans} 
\end{array}
\end{equation}
where $k^n(x)$ are parameters for the general coordinate 
transformations and $s(x)$ is a scaling parameter for the 
Weyl transformation. 

Here it is worth to mention about some algebraic structures of 
the symmetry.  
The first is the reducibility of the symmetry.  
The system is on-shell reducible 
because the gauge transformations \rf{gcs_trans}
have on-shell invariance under the following transformations of 
the gauge parameters with a reducible parameter $w'(x)$, 
\begin{equation}
\begin{array}{rcl}
\delta' u^I
&=& w' \phi^I, 
\\
\delta' \tilde{w}^m 
&=& \displaystyle 
\frac{\ep^{mn}}{\sqrt{-g}} \del_n w'.  
\label{reducibility_condition} 
\end{array}
\end{equation}
Since the transformations (\ref{reducibility_condition}) are not 
reducible anymore, 
the action (\ref{LagrangianBosonicGravity}) is called a first-stage 
reducible system.
The on-shell reducibility is 
the characteristic feature of the gauge symmetry (\ref{gcs_trans}) 
for the generalized Chern-Simons action and the quantization of such a system 
has been discussed in the previous works~\cite{kostu}.  
The second is that the gauge algebra is open. 
This means that the gauge algebra closes only when 
the equations of motion are satisfied.  
Actually, a direct calculation of the commutator of 
two gauge transformations on $\tilde{B}^{mI}(x)$ leads to 
\begin{eqnarray*}
[\delta_1, \delta_2]\tilde{B}^{mI} 
&=& \cdots -(v'_1v_2-v'_2v_1)\frac{\ep^{mn}}{\sqrt{-g}}\del_n\phi^I,  
\end{eqnarray*}
where the dots $(\cdots)$ contain terms of the 
usual ``structure constants'' of the gauge algebra. 
From the points of view of these structures of the gauge symmetry  
we adopt the Batalin-Fradkin-Vilkovisky formulation~\cite{bv, bfv} 
which allows us to deal with reducible and open gauge symmetries 
to obtain covariant gauge-fixed theories. 

%%%%%
\subsection{Semiclassical aspects}

Before getting into the quantization of the system, 
we present semiclassical aspects of the action \rf{LagrangianBosonicGravity}
by eliminating gauge fields through their equations of motion. 
Indeed, this manipulation might be helpful to understand the heart 
of the model. 

First, equations of motion for the fields 
$\tilde{B}^{mI}(x)$ and $\tilde{C}(x)$ give constraints 
\begin{equation}
\begin{array}{rcl}
\del_m\phi_I &=&0, \\
\phi^I\phi_I &=&0.
\end{array}
\end{equation}
The nontrivial solution for these constraints is possible 
if the background spacetime metric includes two 
time-like signatures (\ref{FlatmetricDefinitionGravity}). 
In the light-cone notation\footnote{
%%%%% footnote %%%%%
We use a convention of the light-cone coordinates for the 
background spacetime as 
$x^I = (x^\mu, x^{\hat{+}}, x^{\hat{-}})$ where 
$x^{\hat{\pm}}=\frac{1}{\sqrt{2}}(x^{\hat{0}}\pm x^{\hat{1}})$ and 
the index $\mu$ runs through $0, 1, \dots, D-3$.  
%%%%%%%%%%
}, 
one of the interesting solutions, 
which is naturally related with the usual string action,
is 
$\phi^{\hat{-}}(x)=\phi^\mu(x)=0$ and 
$\phi^{\hat{+}}(x)=\mbox{const.}$. 
By substituting this solution into the classical action 
(\ref{LagrangianBosonicGravity}), the action $S$ becomes 
\begin{equation}
\label{LagrangianBosonicGravityLightcone}
S = \int\! \dd^2 x \, \sqrt{-g} \, 
    \bigg(-\frac{1}{2}g^{mn}\del_m\xi^I\del_n\xi_I  
           -\widetilde{A}^m \phi^{\hat{+}} \del_m \xi^{\hat{-}}
    \,\bigg).
\end{equation}
In the action \rf{LagrangianBosonicGravityLightcone} 
a relation $\del_m \xi^{\hat{-}}(x) = 0$ is given  
by the equation of motion for $\tilde{A}^m(x)$. 
Then, the final form of the action becomes the usual string action 
\begin{equation}
S = \int\! \dd^2 x \, \sqrt{-g} \, 
    \bigg(-\frac{1}{2}g^{mn}\del_m\xi^\mu\del_n\xi_\mu 
    \,\bigg).
\label{string_action}
\end{equation}
Thus, the string action (\ref{string_action}) is regarded as 
a gauge-fixed version of the action (\ref{LagrangianBosonicGravity}). 
The scalar fields $\phi^I(x)$ play an important role 
for the covariant formulation of the {\UvUa} string model 
in the background spacetime which involves 
two time-like coordinates.\footnote{
%%%%% footnote %%%%%
Using this method, 
the noncovariant quantization of the models with an extra time coordinate 
was done in \cite{bdkm, dg}. 
Their models are similar to our model, 
but do not contain the {\UvUa} gauge symmetry. 
%%%%%
} 

From the above manipulation it is suggested 
that the critical dimension of the background spacetime 
is defined as $D-3=25$, {\it i.e.} $D=28$. 
However the dimensions $D$ should be determined in the quantum analysis 
as we will investigate on this paper. 
We also want to emphasize that the quantization will be 
carried out with preserving $D$-dimensional covariance. 
 
%%%%%
\subsection{Preparation for the quantization}

In order to carry out the quantization of the model smoothly, 
we here introduce new $D$ scalar fields $\bar{\phi}^I(x)$ by 
replacing $\tilde{B}^{mI}(x)$ as 
$$
\tilde{B}^{mI} \longrightarrow 
\tilde{B}^{mI} - g^{mn} \del_n \bar{\phi}^I.
$$
Because of the above replacement, 
a new gauge symmetry with a gauge parameter $u'^I(x)$, 
\begin{equation}
\begin{array}{rcl}
\delta \tilde{B}^{mI} &=& g^{mn} \del_n u'^I, \\
\delta \bar{\phi}^I   &=& u'^I, 
\end{array}
\label{u'_trans}
\end{equation}
appears. 
Then, the action \rf{LagrangianBosonicGravity} is modified to 
\begin{eqnarray}
S = \int\! \dd^2 x \, \sqrt{-g} \, 
    \bigg(&-& \frac{1}{2} g^{mn} \partial_m \xi^I \partial_n \xi_I 
          -  g^{mn} \partial_m \bar{\phi}^I \partial_n \phi_I \nonumber \\ 
         &+& \tilde{A}^m \phi_I \partial_m \xi^I 
          +  \tilde{B}^{mI} \partial_m \phi_I
          -  \frac{1}{2} \tilde{C} \phi^I \phi_I
    \,\bigg).
\label{classical_action_02}
\end{eqnarray}
Together with the gauge symmetry (\ref{gcs_trans}) 
the new gauge symmetry (\ref{u'_trans}) constructs another {\UvUa} 
gauge symmetry on the gauge fields $\tilde{B}^{mI}(x)$. 
In particular, these {\UvUa} gauge symmetries turn out to be 
helpful for the covariant quantizations. 

In addition to the gauge symmetry 
(\ref{u1u1})-(\ref{general_coordinate_Weyl_trans}) and (\ref{u'_trans}), 
the action (\ref{classical_action_02}) is also invariant 
under the following global transformations,  
\begin{eqnarray}
\delta \xi^I  
&=& \omega^I{}_J \xi^J + a^I, 
\nonumber\\
\delta \tilde{A}^m  
&=& r \tilde{A}^m + \sum_{i=1}^{2g} \alpha_i h^m_{(i)}, 
\nonumber\\
\delta \phi^I 
&=& - r \phi^I + \omega^I{}_J \phi^J, 
\nonumber \\
\delta \bar{\phi}^I 
&=& r \bar{\phi}^I + \omega^I{}_J \bar{\phi}^J, 
\label{global_trans} \\
\delta \tilde{B}^{mI} 
&=& r \tilde{B}^{mI} + \omega^I{}_J \tilde{B}^{mJ}  
    + \sum_{i=1}^{2g} ( \beta_i^I + \alpha_i \xi^I) h^m_{(i)}, 
\nonumber \\
\delta \tilde{C} 
&=& 2 r \tilde{C}, 
\nonumber \\
\delta g_{mn} 
&=& 0.
\nonumber 
\end{eqnarray}
In the transformation (\ref{global_trans}) 
the parameters $\omega_{IJ} = -\omega_{JI}$, 
$a^I$ and $r$ are global parameters for the $D$-dimensional Lorentz
transformation, the translation and the scale transformation, 
respectively.  
The functions $h_{(i)}^m(x)$ 
are harmonic functions which satisfy 
$\nabla_m h_{(i)}^m(x) = \ep_{mn} \nabla^mh_{(i)}^n(x) = 0$ 
($i$ $=$ $1$, $2$, $\dots$, $2g$; $g=$ genus of two-dimensional 
spacetime) and $\alpha_i$ and $\beta_i^I$ are global parameters. 

%%%%%
\section{Covariant quantization in the Lagrangian formulation}
\setcounter{equation}{0}
\setcounter{footnote}{0}
  
In this section we consider the covariant quantization of the 
action (\ref{classical_action_02}). 
As we explained in the previous section, 
the action has first-stage reducible and open gauge symmetries. 
In order to quantize the action we adopt the 
field-antifield formulation {\it \'a la} Batalin-Vilkovisky.  

In the construction of Batalin-Vilkovisky formulation~\cite{bv}, 
ghost and ghost for ghost 
fields according to the reducibility condition and corresponding
each antifields are introduced. 
The Grassmann parities of the antifields are opposite to 
those of the corresponding fields. 
If a field has ghost number $n$, its antifield has ghost number $-n-1$. 
We denote a set of fields and their antifields by $\Phi^A(x)$ and
$\Phi^*_A(x)$, respectively,   
\begin{eqnarray*}
\Phi^A(x) &=& 
\Big(\varphi^i(x), {\cal C}_0^{a_0}(x), {\cal C}_1^{a_1}(x), 
\ldots, {\cal C}_N^{a_N}(x)
\Big), \\
\Phi^*_A(x) &=& 
\Big(\varphi^*_i(x), {\cal C}^*_{0, a_0}(x), {\cal C}^*_{1, a_1}(x), 
\ldots, {\cal C}^*_{N, a_N}(x)
\Big). 
\end{eqnarray*}
The fields $\varphi^i(x)$ are classical fields, on the other hand, 
the fields ${\cal C}_n^{a_n}(x)$ [$n=0$, $1$, $\ldots$, $N$] 
are ghost and ghost for ghost fields 
corresponding to $N$-th reducible conditions. 
The classical fields $\varphi^i(x)$ and the ghost fields 
${\cal C}_n^{a_n}(x)$ have the ghost number $0$ and $n+1$, respectively.    
Then a minimal action $S_{\rm min}(\Phi, \Phi^*)$ 
is determined by solving the following master equation, 
\begin{equation}
\Big( S_{\rm min}(\Phi, \Phi^*),S_{\rm min}(\Phi, \Phi^*) \Big) =0, 
\label{bv_master_eq}
\end{equation}
with the boundary conditions
\begin{subequations}
\begin{eqnarray}
S_{\rm min}(\Phi, \Phi^*)\bigg|_{\Phi^*=0}
&=& S_{\rm classical}(\varphi),  \\
\frac{\delta_{\rm L}\delta_{\rm R} S_{\rm min}(\Phi, \Phi^*)}
     {\delta{\cal C}^{a_n}_n\delta{\cal C}^*_{n-1, a_{n-1}}}
\bigg|_{\Phi^*=0}
&=& R^{a_{n-1}}_{n, a_{n}}(\Phi), \qquad (n=0, 1, \dots, N).
\label{boundary_condition}
\end{eqnarray}
\end{subequations}

\vspace*{-5mm}
\noindent
Here the antibracket is defined by
\begin{equation}\label{BRSTantibracket}
(X, Y) \equiv 
\frac{\delta_{\rm R}X}{\delta\Phi_A^*}\frac{\delta_{\rm L}Y}{\delta\Phi^A} -
\frac{\delta_{\rm R}X}{\delta\Phi^A}\frac{\delta_{\rm L}Y}{\delta\Phi_A^*}. 
\end{equation}
In this notation, 
${\cal C}^*_{-1, a_{-1}}(x) \equiv \varphi_i^*(x)$ are 
the antifields of the 
classical fields $\varphi^i(x)$. 
The terms $R^{a_{-1}}_{0, a_0}(\Phi)$ and $R^{a_{n-1}}_{n, a_n}(\Phi)$ 
represent the gauge transformations  
and the $n$-th reducibility transformations, respectively. 
The master equation is solved order by order with respect to 
the ghost number.  
The BRST transformations of fields and antifields are given by 
\begin{equation}
s \Phi^A = \Big(S_{\rm min}, \Phi^A\Big), \qquad
s \Phi_A^* = \Big(S_{\rm min}, \Phi_A^*\Big).
\label{bv_brst}
\end{equation}
Eqs.\ (\ref{bv_master_eq}) and (\ref{bv_brst}) assure that the BRST
transformation is nilpotent and the minimal action is invariant 
under the BRST transformation\footnote{
%%%%% footnote %%%%%
Our convention for the Leibniz rule of the BRST operation is given by 
$s(XY)=(sX)Y+(-)^{|X|}X(sY)$, where $|X|$ is a Grassmann parity of 
field $X$.
%%%%%
}. 

Now let us consider to construct the minimal action 
$S_{\rm min}(\Phi, \Phi^*)$ of the model. 
For simplicity of calculation we first redefine field variables as   
$\hat{g}^{mn}(x) \equiv \sqrt{-g(x)}\,g^{mn}(x)$, 
$\hat{A}^{m}(x) \equiv \sqrt{-g(x)}\,\tilde{A}^{m}(x)$, 
$\hat{B}^{mI}(x) \equiv \sqrt{-g(x)}\,\tilde{B}^{mI}(x)$ and  
$\hat{C}(x) \equiv \sqrt{-g(x)}\,\tilde{C}(x)$.  
Using these new field variables, the classical action 
(\ref{classical_action_02}) 
is rewritten as 
\begin{eqnarray}
S_{\rm classical}
= \int\! \dd^2 x \, 
\bigg(&-&\half \hat{g}^{mn} \prt_m \xi^I \prt_n \xi_I 
       - \hat{g}^{mn} \prt_m \bar{\phi}^I \prt_n \phi_I \nonumber\\
      &+&\hat{A}^m \phi_I \prt_m \xi^I
       + \hat{B}^{mI} \prt_m \phi_I 
       - \half \hat{C} \phi^I \phi_I 
       + ( \hat{g} + 1 ) \hZ \, 
\bigg),
\label{classical_action_03}
\end{eqnarray}
where the scalar density field $\hZ(x)$ is a multiplier 
whose equation of motion compensates 
$\hat{g}(x) \equiv \det\hat{g}^{mn}(x) = -1$~\cite{ko}.
We also redefine the gauge transformations 
(\ref{u1u1})-(\ref{general_coordinate_Weyl_trans}) and 
(\ref{u'_trans}) in terms of these new field variables, 
\begin{eqnarray}
\delta \xi^I &=& k^n \prt_n \xi^I + v'\phi^I, 
\nonumber \\
\delta \hat{A}^m &=&   \prt_n ( k^n \hat{A}^m )
                     - \prt_n k^m \hat{A}^n
                     + \ep^{mn} \del_n v 
                     + \hat{g}^{mn} \del_n v', 
\nonumber \\
\delta \phi^I &=&  k^n \prt_n \phi^I, 
\nonumber \\
\delta \bar{\phi}^I &=& k^n \del_n \bar{\phi}^I + u'^I, 
\nonumber \\
\delta \hat{B}^{mI} &=&   \prt_n ( k^n \hat{B}^{mI} )
                        - \prt_n k^m \hat{B}^{nI} 
                        + \ep^{mn} \prt_n u^I
                        + \hat{g}^{mn} \prt_n u'^I
\label{gauge_trans} \\
                     && - \, ( \ep^{mn} v - \hat{g}^{mn} v' ) \prt_n \xi^I
                        - \hat{w}^m \phi^I,
\nonumber \\
\delta\hat{C} &=&   \prt_n ( k^n \hat{C} ) 
                  + \prt_n \hat{w}^n
                  + \prt_n v' \hat{A}^n 
                  - v' \prt_n \hat{A}^n,
\nonumber \\
\delta\hat{g}^{mn} &=& \prt_l (k^l \hat{g}^{mn}) 
                       - \prt_l k^m \hat{g}^{ln} 
                       - \prt_l k^n \hat{g}^{ml},
\nonumber \\
\delta{\hZ} &=& \partial_n(k^n \hZ), 
\nonumber 
\end{eqnarray}
where we denote $\hat{w}^m(x) \equiv \sqrt{-g(x)}\,\tilde{w}^m(x)$. 
The gauge transformation of the multiplier field $\hZ(x)$ is required  
to keep the action (\ref{classical_action_03}) invariant 
under the general coordinate transformations. 
The reducibility condition (\ref{reducibility_condition}) is expressed by 
\begin{equation}
\begin{array}{rcl}
\delta' u^I &=& w' \phi^I, 
\\ 
\delta' \hat{w}^m &=& \ep^{mn} \prt_n w'.  
\end{array} 
\label{reducibility_condition_02}
\end{equation}

The classical fields $\varphi^i(x)$ consist of 
$\xi^I(x)$, $\phi^I(x)$, $\bar{\phi}^I(x)$, 
$\hat{A}^m(x)$, $\hat{B}^{mI}(x)$, $\hat{C}(x)$, 
$\hat{g}^{mn}(x)$ and $\hZ(x)$. 
Here we introduce the ghost fields 
$a(x)$, $a'(x)$, $b^I(x)$, $b'^I(x)$, $\hat{c}^m(x)$ and $d^m(x)$ 
corresponding to the gauge parameters 
$v(x)$, $v'(x)$, $u^I(x)$, $u'^I(x)$, $\hat{w}^m(x)$ and $k^m(x)$  
and a ghost for ghost field $f(x)$ to the reducible parameter $w'(x)$. 
The ghost fields and the ghost for ghost field are 
fermionic and bosonic, respectively. 
Since the {\UvUa} model is a first-stage reducible system, 
the boundary conditions (\ref{boundary_condition}) with $n = 0$, $1$ 
correspond to 
the gauge transformations (\ref{gauge_trans}) and 
the reducibility conditions (\ref{reducibility_condition_02}), 
respectively. 
It is straightforward to solve the master equation  
perturbatively in the order of antifields~\cite{ht, gps}, 
\newcommand{\spaceSmin}{\phantom{\int\! \dd^2 x \,\, \{}}
\begin{eqnarray}
S_{\rm min} 
  &=& S_{\rm classical} \nonumber \\
  &+& \int\! \dd^2 x \, \bigg\{ - \xi^*_I \big(\, 
            d^n \prt_n \xi^I + a' \phi^I \,\big) \nonumber\\
  & &\spaceSmin - \,\hat{A}^*_m \Big(\, 
            \prt_n (d^n \hat{A}^m) - \prt_n d^m \hat{A}^n 
          + \ep^{mn} \prt_n a + \hat{g}^{mn} \prt_n a'
                \,\Big) \nonumber\\
  & &\spaceSmin - \, \phi^{*}_I \big(\, 
            d^n \prt_n \phi^I \,\big) \nonumber\\
  & &\spaceSmin - \, \bar{\phi}^*_I \big(\, 
            d^n \prt_n \bar{\phi}^I + b'^I \,\big)\nonumber\\
  & &\spaceSmin - \,\hat{B}^*_{mI}\Big(\, 
            \prt_n (d^n\hat{B}^{mI})
          - \prt_n d^m \hat{B}^{nI}
          + \ep^{mn} \prt_n b^I
          + \hat{g}^{mn} \prt_n b'^I 
                \nonumber\\
  & &\spaceSmin \phantom{- \,\hat{B}^*_{mI}\Big(\,}
          - \, ( \ep^{mn} a - \hat{g}^{mn} a' ) \prt_n \xi^I
          - \hat{c}^m \phi^I 
          + \half ( f + a a' ) \ep^{mn} \hat{B}_n^{*I} 
                \,\Big) \nonumber\\
  & &\spaceSmin - \,\hat{C}^*  \Big(\, 
            \prt_n (d^n \hat{C}) 
          + \prt_n \hat{c}^n 
          + \prt_n a' \hat{A}^n 
          - a' \prt_n \hat{A}^n 
                \,\Big) \nonumber\\
  & &\spaceSmin - \, \half \, \hat{g}^*_{mn}\Big(\, 
            \prt_k (d^k \hat{g}^{mn}) 
          - \prt_k d^m \hat{g}^{kn} 
          - \prt_k d^n \hat{g}^{mk} \,\Big) \nonumber\\
  & &\spaceSmin - \,\hZ^* \Big(\, \prt_n (d^n \hZ) \,\Big) \nonumber\\
  & &\spaceSmin 
          + a^* \big(\, d^n \prt_n a \,\big) 
          + \,a'^* \big(\, d^n \prt_n a' \,\big) 
          + b^*_I \big(\, d^n \prt_n b^I + f \phi^I \,\big) 
          + b'^*_I \big(\, d^n \prt_n b'^I \,\big) \nonumber\\
  & &\spaceSmin +\,\hat{c}^*_m \Big( 
            \prt_n(d^n \hat{c}^m) - \prt_n d^m \hat{c}^n  
          + ( \ep^{mn} a - \hat{g}^{mn} a' ) \prt_n a'
          + \ep^{mn} \prt_n f 
                \,\Big) \nonumber\\
  & &\spaceSmin + \, d^*_m \big(\, d^n \prt_n d^m \,\big) \nonumber\\
  & &\spaceSmin - \, f^* \big(\, d^n \prt_n f \,\big)
\, \bigg\}.
\end{eqnarray}

The gauge degrees of freedom are fixed by introducing 
a nonminimal action which must be added to the minimal one 
and choosing a suitable gauge-fixing fermion.  
We here choose the orthonormal gauge condition 
$\hat{g}^{mn}(x)=\eta^{mn}$ 
for the world-sheet metric. 
The {\UvUa} gauge parameters $v(x)$, $v'(x)$ 
and the global parameter $\alpha_i$ 
make us possible to choose the gauge $\hat{A}^m(x)=0$. 
In the same way, we can choose the gauge $\hat{B}^{mI}(x)=0$ 
by using the parameters $u^I(x)$, $u'^I(x)$ and $\beta^I_i$. 
We also fix the gauge $\hat{C}(x)=\hat{C}_0$, 
where $\hat{C}_0$ is a constant parameter, 
by using the gauge parameter $\hat{w}^m(x)$. 
In addition to these gauge fixing procedure, 
we also impose the condition 
$\partial_m ( \hat{g}^{mn}(x) \ep_{nk} \hat{c}^k(x))=0$ 
to fix the residual gauge degrees of freedom 
from the reducibility condition. 
In order to adopt all of these gauge fixing conditions, 
we introduce the nonminimal action $S_{\rm nonmin}$, 
\begin{equation}
S_{\rm nonmin} = \int\! \dd^2 x \, \bigg(\,
        \ep^{mn} \hat{a}_m^* Z^a_n + \ep^{mn} \hat{b}_m^{*I} Z^b_{nI}
      + \hat{c}^* Z^c + \half \hd^{*mn} Z^d_{mn} - \bar{f}^* c' 
            \,\bigg), 
\label{non_minimal}
\end{equation}
and the gauge-fixing fermion $\Psi$, 
\begin{equation}
\Psi = \int\! \dd^2 x \, \bigg(\,
                   \ep_{mn} \hat{a}^m \hat{A}^n
                 + \ep_{mn} \hat{b}^m_I \hat{B}^{nI}
                 + c ( \hat{C} - \hat{C}_0 )
                 + \hd_{mn} \hat{g}^{mn}
                 + \bar{f} \prt_m ( \hat{g}^{mn} \ep_{nk} \hat{c}^k )
             \,\bigg),
\end{equation}
where we require traceless conditions  
\begin{equation}
\eta^{mn}\hd_{mn} = \eta_{mn}\hd^{*mn} = 
\eta^{mn}Z^d_{mn} = \eta_{mn}Z^{d*mn} = 0. 
\end{equation}
The antighost fields 
$\hat{a}_m(x)$, $\hat{b}_m^I(x)$, $c(x)$, $c'(x)$ and $\hd_{mn}(x)$ 
are fermionic fields, 
and the auxiliary fields 
$Z^a_m(x)$, $Z^b_{mI}(x)$, $Z^c(x)$, $Z^d_{mn}(x)$ and $\bar{f}(x)$ 
are bosonic ones.  
The ghost numbers of the fields are as follows:
$$
  \begin{array}{lll}
    &\bar{f}
    &\hbox{(ghost number $= -2$)} \\
    &\hat{a}^m, \quad \hat{b}^m_I, \quad c, \quad c', \quad \hd_{mn}
    &\hbox{(ghost number $= -1$)} \\
    &\xi^I, \quad \phi^I, \quad \bar{\phi}^I, \quad
     \hat{A}^m, \quad \hat{B}^{mI}, \quad \hat{C}, \quad
     \hat{g}^{mn}, \quad
    &\\
    &Z^a_m, \quad Z^b_{mI}, \quad Z^c, \quad Z^d_{mn}, \quad \hZ
    &\hbox{(ghost number $= 0$)} \\
    &a, \quad a', \quad b^I, \quad b'^I, \quad \hat{c}^m, \quad d^m
    &\hbox{(ghost number $= 1$)} \\
    &f
    &\hbox{(ghost number $= 2$)} \\
  \end{array}
$$

The BRST transformations of 
$\Phi^A(x)$ and $\Phi^*_A(x)$ are given by 
\begin{equation}
s \Phi^A = \Big(S_{\rm min}+S_{\rm nonmin}, \Phi^A\Big), \qquad
s \Phi_A^* = \Big(S_{\rm min}+S_{\rm nonmin}, \Phi_A^*\Big).
\label{bv_brst2}
\end{equation}
Therefore, the BRST transformations of fields $\Phi^A(x)$ are 
\begin{subequations}
\begin{eqnarray}
s \xi^I &=&  d^n \prt_n \xi^I + a' \phi^I, 
\nonumber \\
s \hat{A}^m &=& 
                  \prt_n ( d^n \hat{A}^m ) - \prt_n d^m \hat{A}^n 
                + \ep^{mn} \prt_n a + \hat{g}^{mn} \prt_n a',
\nonumber \\
s \phi^I &=& d^n \prt_n \phi^I, 
\nonumber \\
s \bar{\phi}^I &=&  d^n \prt_n \bar{\phi}^I + b'^I, 
\nonumber \\
s \hat{B}^{mI} &=& 
                  \prt_n ( d^n \hat{B}^{mI} ) - \prt_n d^m \hat{B}^{nI}
                + \ep^{mn} \prt_n b^I + \hat{g}^{mn} \prt_n b'^I
\nonumber \\
&&              
                - \, ( \ep^{mn} a - \hat{g}^{mn} a' ) \prt_n \xi^I
                - \hat{c}^m \phi^I + ( f + a a' ) \ep^{mn} \hat{B}_n^{*I},
\nonumber \\
s \hat{C} &=& 
                  \prt_n ( d^n \hat{C} )
                + \prt_n \hat{c}^n 
                + \prt_n a' \hat{A}^n - a' \prt_n \hat{A}^n,
\nonumber \\
s \hat{g}^{mn} &=& \prt_k ( d^k \hat{g}^{mn} ) 
                - \prt_k d^m \hat{g}^{kn} - \prt_k d^n \hat{g}^{mk},
\label{BRSTPhi1} \\
s \hZ &=& \prt_n ( d^n \hZ ), 
\nonumber \\
s a &=& d^n \prt_n a, 
\nonumber \\
s a' &=& d^n \prt_n a', 
\nonumber \\
s b^I &=& d^n \prt_n b^I + f \phi^I, 
\nonumber \\
s b'^I &=& d^n \prt_n b'^I, 
\nonumber \\
s \hat{c}^m &=& 
                  \prt_n(d^n \hat{c}^m) - \prt_n d^m \hat{c}^n 
                + ( \ep^{mn} a - \hat{g}^{mn} a' ) \prt_n a' 
                + \ep^{mn} \prt_n f,
\nonumber \\
s d^m &=& d^n \prt_n d^m, 
\nonumber \\
s f &=& d^n \prt_n f, 
\nonumber 
\end{eqnarray}
and 
\begin{eqnarray}
s \hat{a}^m     &=&  \ep^{mn} Z^a_n,   
\hspace*{13mm}   s Z^a_m = 0, 
\nonumber \\
s \hat{b}^m_I   &=&  \ep^{mn} Z^b_{nI},   
\hspace*{10.5mm} s Z^b_{mI} = 0, 
\nonumber \\
s c &=&             Z^c,       
\hspace*{21mm}  s Z^c = 0, 
\label{BRSTPhi2} \\
s \hd_{mn} &=&  Z^d_{mn},   
\hspace*{15mm}  s Z^d_{mn} = 0, 
\nonumber \\
s \bar{f} &=&       c',         
\hspace*{24.5mm}   s c' = 0. 
\nonumber 
\end{eqnarray}
\end{subequations}

\vspace*{-8mm}
\noindent
The antifields are eliminated by using equations 
$\Phi^*_A(x) = \delta_{\rm L}\Psi/\delta\Phi^A(x)$. 
Then the gauge-fixed action is given by 
\newcommand{\spaceSgf}{\phantom{\int\! \dd^2 x \, \big\{\,}}
\begin{eqnarray}
S_{\rm gauge\hbox{-}fixed}
&=& S_{\rm min}+S_{\rm nonmin}
   \Big|_{\Phi^*=\frac{\delta\Psi}{\delta\Phi}} 
\nonumber \\
&=&\int\! \dd^2 x \, \bigg\{
                  - \half \hat{g}^{mn} \prt_m \xi^I \prt_n \xi_I
                  - \hat{g}^{mn} \prt_m \bar{\phi}^I \prt_n \phi_I 
\nonumber \\
& &\spaceSgf     + \, \hat{A}^m \phi_I \prt_m \xi^I
                 + \hat{B}^{mI} \prt_m \phi_I 
                 - \half \hat{C} \phi^I \phi_I
                 + ( \hat{g} + 1 ) \hZ 
\nonumber \\
& &\spaceSgf     + \, \ep_{mk} \hat{a}^k \Big(\, 
                        \prt_n ( d^n \hat{A}^m ) - \prt_n d^m \hat{A}^n
                      + \ep^{mn} \prt_n a + \hat{g}^{mn} \prt_n a'
                   \,\Big)
\nonumber \\
& &\spaceSgf     + \, \ep_{mk} \hat{b}^k_I \Big(\, 
                        \prt_n (d^n \hat{B}^{mI})
                      - \prt_n d^m \hat{B}^{nI}
                      + \ep^{mn} \prt_n b^I
                      + \hat{g}^{mn} \prt_n b'^I
\nonumber \\
& &\spaceSgf   \phantom{+ \, \ep_{mk} \hat{b}^k_I \Big(\,}
                      - \, ( \ep^{mn} a - \hat{g}^{mn} a' ) \prt_n \xi^I
                      - \hat{c}^m \phi^I
                      - \half ( f + a a' ) \hat{b}^{mI} 
                   \,\Big)
\nonumber \\
& &\spaceSgf     - \, c \, \Big(\, 
                        \prt_n (d^n \hat{C}) + \prt_n \hat{c}^n
                      + \prt_n a' \hat{A}^n - a' \prt_n \hat{A}^n
                   \,\Big)
\nonumber \\
& &\spaceSgf     - \, \frac{1}{2} \Big( \hd_{mn}
                 - \prt_m \bar{f} \ep_{nl} \hat{c}^l
                 - \prt_n \bar{f} \ep_{ml} \hat{c}^l
             \Big) \Big( \prt_k (d^k \hat{g}^{mn}) 
                 - \prt_k d^m \hat{g}^{kn} 
                 - \prt_k d^n \hat{g}^{mk} 
             \Big) \nonumber \\
& &\spaceSgf     + \, \ep_{ml} \hat{g}^{lk} \prt_k \bar{f} \Big(
                   \prt_n (d^n \hat{c}^m) - \prt_n d^m \hat{c}^n
                 + ( \ep^{mn} a - \, \hat{g}^{mn} a' ) \prt_n a'
                 + \ep^{mn} \prt_n f 
             \Big) \nonumber \\
& &\spaceSgf     - \, \hat{A}^m Z^a_m - \hat{B}^{mI} Z^b_{mI}
                 + (\hat{C} - \hat{C}_0) Z^c 
                 + \half \hat{g}^{mn} Z^d_{mn}
                 - \prt_m ( \hat{g}^{mn} \ep_{nk} \hat{c}^k) \, c'
         \, \bigg\}
\nonumber \\
&=& \int\! \dd^2 x \, \bigg\{
- \half \hat{g}^{mn} \prt_m \xi^I \prt_n \xi_I 
- \hat{g}^{mn} \prt_m \bar{\phi}^I \prt_n \phi_I 
- \hat{g}^{mn} \prt_m \bar{f} \tinyspace \prt_n f
\nonumber\\&&\spaceSgf
- \Big( \hat{a}^m - \ep^{mn} \prt_n (\bar{f} \tinyspace a) 
                  + \bar{f} \tinyspace \hat{g}^{mn} \prt_n a' 
                  - \hat{b}^m_I \xi^I \Big) 
  \Big( \prt_m a   + \ep_{mk} \hat{g}^{kn} \prt_n a' \Big)
\nonumber\\&&\spaceSgf
- \hat{b}^m_I \Big( \prt_m ( b^I + a \tinyspace \xi^I) + 
                    \ep_{mk} \hat{g}^{kn} \prt_n ( b'^I + a' \xi^I) \Big)
\nonumber\\&&\spaceSgf
- \hat{c}^m \Big( \prt_m c 
       + \ep_{mk} \hat{g}^{kn} \prt_n ( c' - d^n \prt_n \bar{f} ) \Big)
+ \hat{g}^{mn} \hd_{mk} \prt_n d^k
- \half \hat{g}^{mn} d^k \prt_k \hd_{mn}
\nonumber\\&&\spaceSgf
- 2 a \tinyspace \hat{b}^m_I \prt_m \xi^I
+ \ep_{mn} \hat{b}^m_I \hat{c}^n \phi^I 
+ \half (f + a a') \ep_{mn} \hat{b}^m_I \hat{b}^{nI} 
\nonumber\\&&\spaceSgf
- \hat{A}^m \Big( Z^a_m - \phi_I \prt_m \xi^I 
                - \ep_{mn} d^k \prt_k \hat{a}^n 
                - \prt_m d^k \ep_{kn} \hat{a}^n 
                + c \prt_m a' + \prt_m (c \tinyspace a') \Big)
\nonumber\\&&\spaceSgf
- \hat{B}^{mI} \Big( Z^b_{mI} - \prt_m \phi_I 
                - \ep_{mn} d^k \prt_k \hat{b}^n_I 
                - \prt_m d^k \ep_{kn} \hat{b}^n_I \Big)
\nonumber\\&&\spaceSgf
+ \hat{C} \Big( Z^c - \half \phi^I \phi_I - d^n \prt_n c \Big)
- \hat{C}_0 Z^c
\nonumber\\&&\spaceSgf
+ \half \hat{g}^{mn} Z^d_{mn} \, 
+ \, ( \hat{g} + 1 ) 
     \Big( \hZ - \ep^{mn} \prt_m ( \bar{f} \tinyspace a' ) \prt_n a' \Big)
\, \bigg\}.
\label{gauge-fixed_01}
\end{eqnarray}
Here, it should be noted that one can remove a BRST exact term 
$- \hat{C}_0 Z^c(x) = - s ( \hat{C}_0 c(x))$ 
from the above action. 
An influence of the parameter $\hat{C}_0$ disappears 
at the quantum level. 

In order to simplify the form of the action, 
let us redefine some of fields as follows:
\begin{eqnarray*}
Z^a_m - \phi_I \prt_m \xi^I 
      - \ep_{mn} d^k \prt_k \hat{a}^n 
      - \prt_m d^k \ep_{kn} \hat{a}^n 
      + c \prt_m a' + \prt_m (c \tinyspace a')  
  &\rightarrow&  Z^a_m      , 
\\
Z^b_{mI} - \prt_m \phi_I 
         - \ep_{mn} d^k \prt_k \hat{b}^n_I 
         - \prt_m d^k \ep_{kn} \hat{b}^n_I 
  &\rightarrow&  Z^b_{mI}   ,
\\
Z^c - \half \phi^I \phi_I - d^n \prt_n c
  &\rightarrow&  Z^c        ,
\\
\hZ - \ep^{mn} \prt_m ( \bar{f} \tinyspace a' ) \prt_n a'
  &\rightarrow&  \hZ        ,\\
\hat{a}^m - \ep^{mn} \prt_n ( \bar{f} a ) + \bar{f} \, \hat{g}^{mn} \prt_n a'
          - \hat{b}^m_I \xi^I 
  &\rightarrow&  \hat{a}^m  ,
\\
b^I  + a  \tinyspace \xi^I 
  &\rightarrow&  b^I        ,
\\
b'^I + a' \xi^I 
  &\rightarrow&  b'^I       ,
\\
c' - d^n \prt_n \bar{f} 
  &\rightarrow&  c'         .
\end{eqnarray*}
Under these field redefinitions,  
the action \rf{gauge-fixed_01} is modified to 
\begin{eqnarray}
S_{\rm gauge\hbox{-}fixed} &=& \int\! \dd^2 x \, \bigg\{
- \, \half \hat{g}^{mn} \prt_m \xi^I \prt_n \xi_I 
- \hat{g}^{mn} \prt_m \bar{\phi}^I \prt_n \phi_I 
- \hat{g}^{mn} \prt_m \bar{f} \prt_n f
\nonumber\\&&\spaceSgf
- \hat{a}^m   ( \prt_m a   + \ep_{mk} \hat{g}^{kn} \prt_n a' )
- \hat{b}^m_I ( \prt_m b^I + \ep_{mk} \hat{g}^{kn} \prt_n b'^I )
\nonumber\\&&\spaceSgf
- \hat{c}^m   ( \prt_m c   + \ep_{mk} \hat{g}^{kn} \prt_n c' )
+ \hat{g}^{mn} \hd_{mk} \prt_n d^k 
- \frac{1}{2}\hat{g}^{mn}d^k\del_k\bar{d}_{mn} 
\nonumber\\&&\spaceSgf
- 2 a \tinyspace \hat{b}^m_I \prt_m \xi^I
+ \ep_{mn} \hat{b}^m_I \hat{c}^n \phi^I 
+ \half (f + a a') \ep_{mn} \hat{b}^m_I \hat{b}^{nI}
\nonumber\\&&\spaceSgf
- \hat{A}^m Z^a_m - \hat{B}^{mI} Z^b_{mI} + \hat{C} Z^c 
+ \half \hat{g}^{mn} Z^d_{mn}
+ ( \hat{g} + 1 ) \hZ 
\, \bigg\}. 
\label{final_covariant_action}
\end{eqnarray}
The BRST transformations \rf{BRSTPhi1} and \rf{BRSTPhi2} also become 
\begin{eqnarray}
s \xi^I &=&  d^n \prt_n \xi^I + a' \phi^I , 
\nonumber \\
s \phi^I &=& d^n \prt_n \phi^I, 
\nonumber \\
s \bar{\phi}^I &=&  d^n \prt_n \bar{\phi}^I + b'^I - a' \xi^I , 
\nonumber \\
s f &=& d^n \prt_n f , 
\nonumber \\
s \bar{f} &=&  d^n \prt_n \bar{f} + c' ,
\nonumber \\
s a &=& d^n \prt_n a, 
\nonumber \\
s a' &=& d^n \prt_n a', 
\nonumber \\
s b^I &=& d^n \prt_n b^I + ( f - a a' ) \phi^I , 
\nonumber \\
s b'^I &=& d^n \prt_n b'^I, 
\nonumber \\
s c &=&  d^n \prt_n c + \half \phi^I \phi_I + Z^c , 
\nonumber \\
s c' &=& d^n \prt_n c' , 
\nonumber \\
s d^m &=& d^n \prt_n d^m , 
\nonumber \\
s \hat{a}^m &=& 
                  \prt_n ( d^n \hat{a}^m ) - \prt_n d^m \hat{a}^n 
                + \ep^{mn} ( \phi_I \prt_n \xi^I - \prt_n \phi_I \xi^I ) 
                - a' \tinyspace \hat{b}^m_I \phi^I 
\nonumber \\
&&
                - \, ( \ep^{mn} c - \hat{g}^{mn} c' ) \prt_n a'
                - \ep^{mn} \prt_n ( c a' + c' a )
                + \ep^{mn} ( Z^a_n - Z^b_{nI} \xi^I ) ,
\label{final_BRSTPhi} \\
s \hat{b}^m_I &=& 
                    \prt_n ( d^n \hat{b}^m_I ) - \prt_n d^m \hat{b}^n_I
                  + \ep^{mn} \prt_n \phi_I
                  + \ep^{mn} Z^b_{nI} ,
\nonumber \\
s \hat{c}^m &=& 
                  \prt_n(d^n \hat{c}^m) - \prt_n d^m \hat{c}^n 
                + ( \ep^{mn} a - \hat{g}^{mn} a' ) \prt_n a' 
                + \ep^{mn} \prt_n f ,
\nonumber \\
s \hd_{mn} &=&  Z^d_{mn} ,
\nonumber \\
s \hat{A}^m &=& 
                  \prt_n ( d^n \hat{A}^m ) - \prt_n d^m \hat{A}^n 
                + \ep^{mn} \prt_n a + \hat{g}^{mn} \prt_n a' ,
\nonumber \\
s \hat{B}^{mI} &=& 
                  \prt_n ( d^n \hat{B}^{mI} ) - \prt_n d^m \hat{B}^{nI}
                + \ep^{mn} \prt_n b^I + \hat{g}^{mn} \prt_n b'^I
\nonumber \\
&&              
                - \, \ep^{mn} \big( a \prt_n \xi^I + \prt_n ( a \xi^I ) \big)
                - \hat{g}^{mn} \prt_n a' \xi^I 
                - \hat{c}^m \phi^I - ( f + a a' ) \hat{b}^{mI} ,
\nonumber \\
s \hat{C} &=& 
                  \prt_n ( d^n \hat{C} )
                + \prt_n \hat{c}^n 
                + \prt_n a' \hat{A}^n - a' \prt_n \hat{A}^n ,
\nonumber \\
s \hat{g}^{mn} &=& \prt_k ( d^k \hat{g}^{mn} ) 
                - \prt_k d^m \hat{g}^{kn} - \prt_k d^n \hat{g}^{mk} ,
\nonumber \\
s Z^a_m &=& d^n \prt_n Z^a_m + \prt_m d^n Z^a_n 
                + Z^c \prt_m a' + \prt_m ( Z^c a' ) ,
\nonumber \\
s Z^b_{mI} &=& d^n \prt_n Z^b_{mI} + \prt_m d^n Z^b_{nI} ,
\nonumber \\
s Z^c &=& d^n \prt_n Z^c ,
\nonumber \\
s Z^d_{mn} &=& 0, 
\nonumber \\
s \hZ &=& \prt_n ( d^n \hZ ) - \ep^{mn} \prt_m ( c' a' ) \prt_n a'.
\nonumber 
\end{eqnarray}
The action \rf{final_covariant_action} is invariant 
under the nilpotent BRST transformations \rf{final_BRSTPhi}. 

Using equations of motion for the fields 
$Z^a_m(x)$, $Z^b_{mI}(x)$, $Z^c(x)$, $Z^d_{mn}(x)$, $\hZ(x)$, 
$\hat{A}^m(x)$, $\hat{B}^{mI}(x)$, $\hat{C}(x)$ and $\hat{g}^{mn}(x)$,   
thus imposing gauge fixing conditions, we fix fields as 
\begin{equation}
\begin{array}{rcl}
  &&\hat{A}^m = \hat{B}^{mI} = \hat{C} = 0 , \qquad\ 
    \hat{g}^{mn} = \eta^{mn} , \\
  &&\displaystyle 
    Z^a_m = Z^b_{mI} = Z^c = 0 , \qquad
    Z^d_{mn}  =  V_{mn} - \half \eta_{mn} \eta^{kl} V_{kl} , \qquad
    \hZ  =  - \,\frac{1}{4} \eta^{mn} V_{mn} , 
\end{array}
\end{equation}
where we denote 
\begin{eqnarray}
V_{mn}
&=& \half \prt_m \xi^I \prt_n \xi_I
    + \prt_m \bar{\phi}^I \prt_n \phi_I
    + \prt_m \bar{f} \tinyspace \prt_n f
\nonumber\\&&
    + \, \hat{a}^k \ep_{km} \prt_n a'
    + \hat{b}^k_I \ep_{km} \prt_n b'^I
    + \hat{c}^k \ep_{km} \prt_n c'
    - \hd_{mk} \prt_n d^k 
    + \frac{1}{2} d^k \prt_k \hd_{mn}
\nonumber\\&&
    + \, (m \leftrightarrow n).  
\end{eqnarray}
Then, we finally obtain the following gauge-fixed action, 
\begin{eqnarray}
S_{\rm gauge\hbox{-}fixed}
= \int\! \dd^2 x \, \bigg\{
&-& \half \eta^{mn} \prt_m \xi^I \prt_n \xi_I 
  - \eta^{mn} \prt_m \bar{\phi}^I \prt_n \phi_I 
  - \eta^{mn} \prt_m \bar{f} \tinyspace \prt_n f 
\nonumber \\
&-& \hat{a}^m   \big( \prt_m a + \ep_m{}^n \prt_n a' \big) 
  - \hat{b}^m_I \big( \prt_m b^I + \ep_m{}^n \prt_n b'^I \big) 
\nonumber \\
&-& \hat{c}^m \big( \prt_m c + \ep_m{}^n \prt_n c' \big) 
  + \eta^{mn} \hd_{mk} \prt_n d^k 
\nonumber \\
&-& 2 a \tinyspace \hat{b}^m_I \prt_m \xi^I
  + \ep_{mn} \hat{b}^m_I \hat{c}^n \phi^I 
  + \half (f + a a') \ep_{mn} \hat{b}^m_I \hat{b}^{nI}
\, \bigg\} .
\label{gauge-fixed_02}
\end{eqnarray}
The action (\ref{gauge-fixed_02}) is invariant 
under the following on-shell nilpotent BRST transformation, 
in which the antifields and the auxiliary fields are eliminated, 
\begin{eqnarray}
s \xi^I &=&  d^n \prt_n \xi^I + a' \phi^I , 
\nonumber \\
s \phi^I &=& d^n \prt_n \phi^I, 
\nonumber \\
s \bar{\phi}^I &=&  d^n \prt_n \bar{\phi}^I + b'^I - a' \xi^I ,
\nonumber \\
s f &=& d^n \prt_n f , 
\nonumber \\
s \bar{f} &=&  d^n \prt_n \bar{f} + c' , 
\nonumber \\
s a &=&    d^n \prt_n a ,
\nonumber \\
s a' &=&   d^n \prt_n a' ,
\nonumber \\
s b^I &=&   d^n \prt_n b^I + ( f - a a' ) \phi^I ,
\nonumber \\
s b'^I &=&  d^n \prt_n b'^I ,
\label{lagrangian_BRST}\\
s c &=&   d^n \prt_n c + \half \phi^I \phi_I ,
\nonumber \\
s c' &=& d^n \prt_n c' , 
\nonumber \\
s d^m &=& d^n \prt_n d^m , 
\nonumber \\
s \hat{a}^m &=& 
                  \prt_n ( d^n \hat{a}^m ) - \prt_n d^m \hat{a}^n 
                + \ep^{mn} ( \phi_I \prt_n \xi^I - \prt_n \phi_I \xi^I ) 
                - a' \tinyspace \hat{b}^m_I \phi^I 
\nonumber \\
&&              
                - ( \ep^{mn} c - \eta^{mn} c' ) \prt_n a' 
                - \, \ep^{mn} \prt_n ( c \tinyspace a' + c' a ) ,
\nonumber \\
s \hat{b}^m_I &=&   
                  \prt_n ( d^n \hat{b}^m_I ) - \prt_n d^m \hat{b}^n_I
                + \ep^{mn} \prt_n \phi_I ,
\nonumber \\
s \hat{c}^m &=& 
                  \prt_n(d^n \hat{c}^m) - \prt_n d^m \hat{c}^n 
                + ( \ep^{mn} a - \eta^{mn} a' ) \prt_n a' 
                + \ep^{mn} \prt_n f ,
\nonumber \\
s \hd_{mn} &=&  V_{mn} - \half \eta_{mn} \eta^{kl} V_{kl} .
\nonumber 
\end{eqnarray}
We can check the following relations for 
the nilpotency of the BRST transformation, 
\begin{eqnarray*}  
s^2 \hat{c}^m 
&=& - \, 2 \bigg( \frac{\delta_{\rm L}S}{\delta \hd_{mn}} \bigg) 
           a' \prt_n a' ,\\
s^2 \hd_{mn} 
&=& - \, \bigg( \frac{\delta_{\rm L}S}{\delta \hat{c}^m}\bigg)
                    a' \prt_n a'
                    - \bigg(\frac{\delta_{\rm L}S}{\delta \hat{c}^n} \bigg)
                    a' \prt_m a' + \eta_{mn} \eta^{kl}
                    \bigg(\frac{\delta_{\rm L}S}{\delta \hat{c}^k} \bigg)
                    a' \prt_l a'  \\
&&             + \, \eta_{mk} \eta_{nl} 
                    \bigg(\frac{\delta_{\rm L}S}{\delta \hd_{kl}} \bigg)
                    \eta^{pq} V_{pq}, \\
s^2 (\mbox{others}) 
&=& 0 .
\end{eqnarray*}

Now we present a perturbative analysis of the gauge-fixed action.
We would like to investigate the BRST Ward identities at 
the quantum level. 
Then, we find out that nonlocal anomalous terms obtained from 
one-loop calculations vanish by imposing a condition, 
which determines the critical dimension for this string model. 
For the explicit calculation 
it is convenient to introduce light-cone notations on 
the world-sheet\footnote{
%%%%% footnote %%%%%
Our convention of the light-cone coordinates on the world-sheet is  
$x^{\pm}=\frac{1}{\sqrt{2}}(x^0 \pm x^1)$. 
The metric tensor and the Levi-Civit\'a symbol are given by 
$\eta_{++}=\eta_{--}=0, \ \eta_{+-}=\eta_{-+}=-1$ and 
$\varepsilon_{+-}=-\varepsilon_{-+}=-1$, respectively.
%%%%%
}.
Then, the gauge-fixed action (\ref{gauge-fixed_02}) is expressed 
with these notations
\begin{eqnarray}
S_{\rm gauge\hbox{-}fixed}
= \int\! \dd^2 x \, \bigg\{
&& \prt_+ \xi^I \prt_- \xi_I 
+ 2 \prt_+ \bar{\phi}^I \prt_- \phi_I 
+ 2 \prt_+ \bar{f} \, \prt_- f 
\nonumber \\
&& + \, \hat{a}_+ \prt_- a_+ + \hat{a}_- \prt_+ a_- 
   +    \hat{b}_{+I} \prt_- b_+^I + \hat{b}_{-I} \prt_+ b_-^I 
\nonumber \\
&& + \, \hat{c}_+ \prt_- c_+ + \hat{c}_- \prt_+ c_- 
   -    \hd_{++} \prt_- d^+ - \hd_{--} \prt_+ d^- 
\nonumber \\
&& + \, (a_+ + a_-) 
        (\, \hat{b}_{+I} \prt_- \xi^I + \hat{b}_{-I} \prt_+ \xi^I \,)
\nonumber \\
&& + \, \phi^I \hat{b}_{+I} \hat{c}_- 
   -    \phi^I \hat{b}_{-I} \hat{c}_+ 
   + \Big( f + \half a_+ a_- \Big) \, \hat{b}_{+I} \hat{b}_-^I
\, \bigg\}, 
\label{gauge_fixed_03}
\end{eqnarray}
where we denote 
$a_\pm(x)   \equiv  a(x) \mp a'(x)$, 
$b_\pm^I(x) \equiv  b^I(x) \mp b'^I(x)$ and 
$c_\pm(x)   \equiv  c(x) \mp c'(x)$. 
Propagators are derived by taking inverses of bilinear parts in 
the action (\ref{gauge_fixed_03}),  
\begin{eqnarray*}
\langle \xi^{I}(x) \xi^J(y) \rangle_0 
&=& \langle \bar{\phi}^I(x) \phi^J(y) \rangle_0 \\ 
&=& \int\! \frac{\dd^2 p}{i(2\pi)^2}
    \frac{1}{p^2+i\epsilon} \, e^{-ip(x-y)}\eta^{IJ}, \\
\langle \bar{f}(x) f(y) \rangle_0
&=& \int\! \frac{\dd^2 p}{i(2\pi)^2} 
    \frac{1}{p^2+i\epsilon} \, e^{-ip(x-y)}, \\
\langle \hat{a}_{\pm}(x) a_{\pm}(y) \rangle_0 
&=& \langle \hat{c}_{\pm}(x){c}_{\pm}(y) \rangle_0 
= - \langle \hd_{\pm\pm}(x) d^\pm(y) \rangle_0 \\
&=& \int\! \frac{\dd^2 p}{i(2\pi)^2}
    \frac{2ip^{\mp}}{p^2+i\epsilon} \, e^{-ip(x-y)},\\
\langle \hat{b}_{\pm}^I(x) b_{\pm}^J(y) \rangle_0
&=& \int\! \frac{\dd^2 p}{i(2\pi)^2}
    \frac{2ip^{\mp}}{p^2+i\epsilon} \, e^{-ip(x-y)}\eta^{IJ}.
\end{eqnarray*}

Now let us consider the following two-point function,
\begin{equation}
A(p)_{++} \equiv \int\! \frac{\dd^2 x}{i(2\pi)^2}
\langle V_{++}(x) V_{++}(0) \rangle \, e^{ipx}.
\label{v+v+}
\end{equation}
Here we mention that the two-point function (\ref{v+v+}) should vanish 
from the BRST symmetry $V_{++}(x) = s \hd_{++}(x)$.   
Estimating all contributions arising from pairs  
$(\xi^I, \xi_I)$, $(\bar{\phi}^I, \phi_I)$, 
$(\hat{a}_+, a_+)$, $(\hat{b}_{+I}, b_+^I)$, $(\hat{c}_+, c_+)$,
$(\hd_{++}, d^+)$ and $(\bar{f}, f)$ 
we can obtain the following result up to one-loop order, 
\begin{eqnarray}
A(p)_{++}
&=& \frac{1}{48\pi^3} \Big( D+2D-2-2D-2-26+2 \Big) \frac{(p^-)^3}{p^+}
\nonumber \\
&=& \frac{D-28}{48\pi^3} \frac{(p^-)^3}{p^+}.
\end{eqnarray}
In a similar way we obtain 
\begin{equation}
A(p)_{--} = \frac{D-28}{48\pi^3}\frac{(p^+)^3}{p^-}.
\end{equation}
Next we evaluate the other type of the two-point functions
\begin{eqnarray*}
A(p)_{+-}
&\equiv&\int\! \frac{\dd^2 x}{i(2\pi)^2}
   \langle V_{++}(x)V_{--}(0)\rangle \, e^{ipx} 
\nonumber \\
&=&- \, \frac{D-8}{8\pi^2}\int\frac{\dd k^+ \dd k^-}{i(2\pi)^2}
        \frac{k^+k^-}{k^+k^-+i\epsilon}
        \frac{(p-k)^+(p-k)^-}{(p-k)^+(p-k)^-+i\epsilon}. 
\end{eqnarray*}
This two-point function is actually quadratically divergent. 
This divergent, however, will be absorbed adding a suitable local
counter term to the action. 
We conclude then that the BRST anomaly vanishes if and only if 
\begin{equation}
D=28.
\end{equation}
 
%%%%%
\section{Covariant quantization in the Hamiltonian formulation}
\setcounter{equation}{0}
\setcounter{footnote}{0}

In this section we carry out the quantization of 
the classical action (\ref{classical_action_02}) in 
the covariant Hamiltonian formulation given by Batalin, Fradkin and 
Vilkovisky. 
We present that the gauge-fixed action and the BRST transformation 
obtained in this formulation coincide with the result of the 
Lagrangian formulation if we make a proper choice of 
a gauge-fermion and a suitable identification of ghosts and 
ghost momenta. 
We also obtain the BRST charge in this formulation. 

First of all we decompose the world-sheet metric $g_{mn}(x)$ 
by using the following convenient parameterization~\cite{bh},
\begin{equation}
g_{mn} = 
\left( 
\begin{array}{cc}
-N^2 \gamma + N_1^2 \gamma & \ \ N_1 \gamma \\
N_1 \gamma & \gamma \\
\end{array}
\right),
\end{equation}
where $N(x)$ and $N_1(x)$ are 
the rescaled lapse and the shift function, respectively. 
Under these parameterization the factor $\gamma(x)$ 
decouples from Weyl invariant theory. 

According to the ordinary Dirac's procedure, 
we introduce the following canonical momenta defined by 
$P_{\Phi^A}(x) \equiv \delta_{\rm L}S/\delta(\del_0\Phi^A(x))$ 
corresponding to fields $\Phi^A(x)$,
\begin{eqnarray}
P_{\xi}^I &=& \frac{1}{N} \del_0\xi^I 
            - \frac{N_1}{N} \del_1\xi^I
            - A_1\phi^I, 
\nonumber \\
P_{\phi}^I &=& \frac{1}{N} \del_0\bar{\phi}^I
             - \frac{N_1}{N} \del_1\bar{\phi}^I 
             - B_{1}^I, 
\label{canonical_momenta}\\
P_{\bar{\phi}}^I &=& \frac{1}{N} \del_0\phi^I
                     - \frac{N_1}{N} \del_1\phi^I, 
\nonumber
\end{eqnarray}
and 
\begin{equation}
P_N = P_{N_1} 
= P_{A_m} = P_{B_m}^I = P_{C_{01}} =0. 
\label{primary}  
\end{equation}
The relations (\ref{primary}) give primary constraints. 
A consistency check of these primary constraints yields a 
set of secondary constraints
\begin{eqnarray}
&&\frac{1}{2}\Big(P_\xi^IP_{\xi I} + \del_1 \xi^I \del_1 \xi_I
             \Big)=0, \label{1st_class_1}\\
&&P_{\xi}^I \del_1 \xi_I=0, \label{1st_class_2}\\
&&\phi_I \del_1 \xi^I = 0, \label{1st_class_3}\\
&&\phi_I P_{\xi}^I = 0, \label{1st_class_4}\\
&&\del_1 \phi^I = 0, \label{1st_class_5}\\
&&P_{\bar{\phi}}^I = 0, \label{1st_class_6}\\ 
&&\frac{1}{2} \phi^I \phi_I = 0, \label{1st_class_7}
\end{eqnarray}
and these conditions give no other relations. 
The constraints (\ref{1st_class_1}) and (\ref{1st_class_2}) 
correspond to the Virasoro constraints. 
We can easily show that the set of these constraints 
(\ref{primary}) and (\ref{1st_class_1})-(\ref{1st_class_7}) 
is first-class.
Introducing Lagrange multiplier fields $\lambda_i(x)$ 
corresponding to primary constraints (\ref{primary}), 
a total Hamiltonian is given by
\begin{eqnarray}
H = \int\! \dd x^1
&\bigg\{\,& N \Big(\,
          \frac{1}{2} (P_\xi^I + A_1\phi^I)(P_{\xi I} + A_1\phi_I) 
          + \frac{1}{2} \del_1 \xi^I \del_1 \xi_I
\nonumber \\&\phantom{\bigg(\,}&\phantom{N \Big(\,}
          + (P_\phi^I + B_1^I) P_{\bar{\phi}I} 
          + \del_1 \bar{\phi}^I \del_1 \phi_I 
             \Big)
\nonumber \\&\phantom{\bigg(\,}&
+ \, N_1 \Big( (P_{\xi}^I + A_1\phi^I) \del_1 \xi_I 
              + (P_{\phi}^I + B_{1}^I) \del_1 \phi_I
              + P_{\bar{\phi}}^I \del_1 \bar{\phi}_I 
         \Big)
\nonumber \\&\phantom{\bigg(\,}&
- \,  A_0 \phi_I \del_1\xi^I 
      - B_0^I \del_1 \phi_I
      - \frac{1}{2} C_{01} \phi^I\phi_I 
\nonumber \\&\phantom{\bigg(\,}&
+ \, \lambda_N P_N + \lambda_{N_1}P_{N_1}
     + \lambda_{A_m} P_{A_m}
     + \lambda_{B_m^I} P_{B_m}^I
     + \lambda_{C_{01}} P_{C_{01}}
 \bigg\}. 
\label{hamiltonian}
\end{eqnarray}
The total Hamiltonian (\ref{hamiltonian}) is 
weakly vanishing on the constraint surface defined by 
(\ref{primary}) and (\ref{1st_class_1})-(\ref{1st_class_7}). 
The gauge transformations of the canonical momenta defined by 
(\ref{canonical_momenta}) are given by 
\begin{eqnarray*}
\delta P_{\xi}^I 
&=&   \del_1 \Big( k^0 N \del_1 \xi^I
    + (k^1 + k^0 N_1) P_{\xi}^I
           \Big) 
\\
&&  - \, \del_1 v \tinyspace \phi^I 
    - \del_1 \Big( k^0 (A_0 - N_1 A_1) \phi^I
             \Big) 
    + v' P_{\bar{\phi}}^I, 
\\
\delta P_{\phi}^I 
&=& - \, P_{\xi}^I \Big( v' + k^0 N A_1 \Big) 
    + \phi^I \Big( w_1 - A_1 v' - k^0 N A_1^2 + k^0 C_{01}
             \Big) 
\\
&& - \, \del_1 \Big( u^I
              - k^0 N \del_1 \bar{\phi}^I
              - ( k^1 + k^0 N_1 ) P_{\phi}^I 
              + k^0 (B_{0}^I - N_1 B_{1}^I)
            \Big)
\\
&& + \, \Big( v + k^0 (A_0 - N_1 A_1)
     \Big) \del_1 \xi^I, 
\\
\delta P_{\bar{\phi}}^I 
&=&  \del_1 \Big( k^0 N \del_1 \phi^I 
                   + (k^1 + k^0 N_1) P_{\bar{\phi}}^I \Big).
\end{eqnarray*}

In the construction of Batalin-Fradkin-Vilkovisky formulation~\cite{bfv} 
a phase space is extended so as to contain 
ghosts $\eta^A(x)$ and 
their canonically conjugate ghost momenta ${\cal P}_A(x)$  
corresponding to constraints $G_{\eta^A}(x)$.   
Then a nilpotent BRST transformation is constructed and 
a physical phase space is defined as its cohomology 
which is a set of gauge invariant functions on the constraint surface.
The role of the ghost momenta is to exclude functions vanishing 
on the constraint surface from the cohomology and gauge invariant 
functions are removed from the cohomology because of the action of the 
BRST transformation for the ghost.

First of all we separate the variables into dynamical and 
non-dynamical ones. 
By adopting gauge conditions $N(x)=1$, $N_1(x)=0$, $A_m(x)=0$, 
$B_m^I(x)=0$ 
and $C_{01}(x)=-\hat{C}(x)=-\hat{C}_0$ \ $(\mbox{const.})$,  
we have a set of dynamical phase space 
variables $\Big(\xi^I(x), P_{\xi}^J(x)\Big)$, 
$\Big(\phi^I(x), P_{\phi}^J(x)\Big)$ and 
$\Big(\bar{\phi}^I(x), P_{\bar{\phi}}^J(x)\Big)$ 
with the first-class constraints
(\ref{1st_class_1})-(\ref{1st_class_7}).

Here we rearrange the first-class 
constraints (\ref{1st_class_1})-(\ref{1st_class_7}) 
into the following forms, 
\begin{eqnarray}
G_{\eta_0}
&=& \frac{1}{2} \Big(P^I_\xi P_{\xi I}
                     + \del_1\xi^I \del_1\xi_I \Big) 
  + P^I_{\bar{\phi}} P_{\phi I} 
  + \del_1 \bar{\phi}^I \del_1\phi_I ,  
\nonumber \\
G_{\eta_1}
&=& P_{\xi}^I\del_1\xi_I
   +P_{\bar{\phi}}^I\del_1\bar{\phi}_I
   +P_{\phi}^I\del_1\phi_I, 
\nonumber \\
G_\eta
&=& \phi_I \del_1 \xi^I, 
\nonumber \\
G_{\eta'}
&=& \phi_I P_{\xi}^I, 
\label{1st_1_1}\\
G_{\eta^I}
&=& \del_1 \phi^I, 
\nonumber \\
G_{\eta'^I}
&=& P_{\bar{\phi}}^I, 
\nonumber \\
G_{\bar{\eta}}
&=& \frac{1}{2} \phi^I \phi_I, 
\nonumber 
\end{eqnarray}
and introduce corresponding canonically conjugate pairs 
of ghosts and ghost momenta 
$\Big(\eta_0(x), {\cal P}_0(x)\Big)$, \hspace*{-1.5mm}
$\Big(\eta_1(x), {\cal P}_1(x)\Big)$, \hspace*{-1.5mm}
$\Big(\eta(x), {\cal P}(x)\Big)$, \hspace*{-1.5mm}
$\Big(\eta'(x), {\cal P}'(x)\Big)$, \hspace*{-1.5mm}
$\Big(\eta^I(x), {\cal P}^J(x)\Big)$, \hspace*{-1.5mm}
$\Big(\eta'^I(x), {\cal P}'^J(x)\Big)$ \hspace*{-1.5mm}
and $\Big(\bar{\eta}(x), \bar{{\cal P}}(x)\Big)$. 
Though the rearrangement of the constraints is not inevitable, 
it turns out that to choose these combinations of the constraints 
is the simplest way to lead to the gauge-fixed 
action (\ref{gauge-fixed_02}) in the covariant Hamiltonian 
formulation. 

As we explained in the previous section, the model has 
the reducible symmetry. 
Indeed the constraints $G_{\eta^I}(x)$ and $G_{\bar{\eta}}(x)$ 
are not independent due to the following relation, 
\begin{equation}
G_{\eta''} \equiv \del_1 G_{\bar{\eta}} - \phi_I G_{\eta^I} = 0. 
\end{equation}
Therefore 
it is necessary to introduce one more Grassmann even 
ghost $\eta''(x)$ and its momentum ${\cal P}''(x)$ corresponding to this  
reducibility condition. 

After the step by step construction according to the systematic 
procedure~\cite{ht}, we obtain the following BRST transformations
in the extended phase space, 
\begin{eqnarray}
s\xi^I
&=&- \, P^I_\xi \eta_0
   - \del_1 \xi^I \eta_1
   - \phi^I \eta'
   - {\cal P}^I \eta_0 \eta
   + {\cal P}'^I \eta_0 \eta' ,
\nonumber \\
sP_{\xi}^I
&=&- \, \del_1 (\del_1 \xi^I \eta_0)
   - \del_1 (P_{\xi}^I\eta_1)
   - \del_1 (\phi^I\eta)
   + \del_1 ({\cal P}'^I\eta_0\eta)
   - \del_1 ({\cal P}^I\eta_0\eta') ,
\nonumber \\
s\phi^I
&=&- \, P^I_{\bar{\phi}} \eta_0
   - \del_1 \phi^I\eta_1, 
\nonumber \\
sP_{\phi}^I
&=&- \, \del_1 (\del_1 \bar{\phi}^I \eta_0)
   - \del_1 (P_{\phi}^I \eta_1)
   + P_{\xi}^I \eta'
   + \del_1 \xi^I \eta
   + \phi^I \bar{\eta}
   - \del_1 \eta^I
   - {\cal P}'^I \eta_0 \bar{\eta}
   - {\cal P}^I \eta'', 
\nonumber \\
s\bar{\phi}^I
&=&- \, P^I_\phi \eta_0
   - \del_1 \bar{\phi}^I \eta_1
   - \eta'^I, 
\nonumber \\
sP_{\bar{\phi}}^I
&=&- \, \del_1 (\del_1 \phi^I \eta_0)
   - \del_1 (P_{\bar{\phi}}^I \eta_1), 
\nonumber \\
s\eta_0
&=&- \, \eta_0 \del_1 \eta_1
   - \eta_1 \del_1 \eta_0, 
\nonumber \\
s{\cal P}_0
&=&- \, \frac{1}{2} P^I_\xi P_{\xi I}
   - \frac{1}{2} \del_1 \xi^I \del_1\xi_I
   - P^I_{\bar{\phi}} P_{\phi I}
   - \del_1 \bar{\phi}^I \del_1 \phi_I
\nonumber \\
&&+ \, P_\xi^I {\cal P}_I \eta 
  - P^I_\xi {\cal P}'_I \eta'
  - \del_1 \xi^I {\cal P}'_I \eta
  + \del_1 \xi^I {\cal P}_I \eta' 
  - \phi^I {\cal P}'_I \bar{\eta}
\nonumber \\
&& + \, {\cal P}_0 \del_1 \eta_1
   + \del_1 ({\cal P}_0 \eta_1)
   + {\cal P}_1 \del_1 \eta_0
   + \del_1 ({\cal P}_1 \eta_0)
   + {\cal P}' \del_1 \eta
   + {\cal P} \del_1 \eta'  
   + {\cal P}'^I \del_1 \eta_I 
   + {\cal P}^I \del_1 \eta'_I 
\nonumber \\
&& - \, {\cal P}^I {\cal P}'_I \eta''
   - {\cal P}'' \eta \del_1 \eta
   - {\cal P}'' \eta' \del_1 \eta'
   + {\cal P}'' \bar{\eta} \del_1 \eta_0
   + \del_1 ({\cal P}'' \bar{\eta} \eta_0),
\nonumber \\
s\eta_1
&=&- \, \eta_0 \del_1 \eta_0
   - \eta_1 \del_1 \eta_1, 
\nonumber \\
s{\cal P}_1 
&=& - \, P_\xi^I \del_1 \xi_I
    - P_\phi^I \del_1 \phi_I 
    - P_{\bar{\phi}}^I \del_1 \bar{\phi}_I 
\nonumber \\
&& + \, {\cal P}_0 \del_1 \eta_0 
   + \del_1 ({\cal P}_0 \eta_0) 
   + {\cal P}_1 \del_1 \eta_1
   + \del_1 ({\cal P}_1 \eta_1)
\nonumber \\
&& + \, {\cal P} \del_1 \eta
   + {\cal P}' \del_1 \eta'
   + {\cal P}^I \del_1 \eta_I
   + {\cal P}'^I \del_1 \eta'_I
   - \del_1 \bar{{\cal P}} \bar{\eta}
   - {\cal P}'' \del_1 \eta'', 
\label{hamiltonian_BRST}\\
s\eta
&=&- \, \eta_0 \del_1 \eta'
   - \eta_1 \del_1 \eta, 
\nonumber \\
s{\cal P} 
&=& - \, \phi_I \del_1 \xi^I
    - P_\xi^I {\cal P}_I \eta_0 
    + \del_1 \xi^I {\cal P}'_I \eta_0 
\nonumber \\
&&  + \, \bar{{\cal P}} \del_1 \eta' 
    + \del_1 (\bar{{\cal P}} \eta')
    + \del_1 ({\cal P}' \eta_0)
    + \del_1 ({\cal P} \eta_1)
    + {\cal P}'' \eta_0 \del_1 \eta 
    + \del_1 ({\cal P}'' \eta_0 \eta),
\nonumber \\
s\eta' 
&=&- \, \eta_0 \del_1 \eta
   - \eta_1 \del_1 \eta', 
\nonumber \\
s{\cal P}' 
&=& - \, \phi_I P_{\xi}^I 
    + P_\xi^I {\cal P}'_I \eta_0
    - \del_1 \xi^I {\cal P}_I \eta_0 
\nonumber \\
&&  + \, \bar{\cal P} \del_1 \eta 
    + \del_1 (\bar{\cal P} \eta)
    + \del_1 ({\cal P} \eta_0)
    + \del_1 ({\cal P}' \eta_1)
    + {\cal P}'' \eta_0 \del_1 \eta' 
    + \del_1 ({\cal P}''\eta_0\eta'), 
\nonumber \\
s\eta^I
&=&- \, \eta_0 \del_1 \eta'^I
   - \eta_1 \del_1 \eta^I
   - \del_1 \xi^I \eta_0 \eta'
   - P_\xi^I \eta_0 \eta
   - {\cal P}'^I \eta_0 \eta''
   + \phi^I \eta'', 
\nonumber \\
s{\cal P}^I 
&=& - \, \del_1 \phi^I 
    + \del_1 ({\cal P}'^I \eta_0)
    + \del_1 ({\cal P}^I \eta_1),
\nonumber \\
s\eta'^I
&=&- \, \eta_0 \del_1 \eta^I
   - \eta_1 \del_1 \eta'^I
   + \phi^I \eta_0 \bar{\eta}
   + P^I_\xi \eta_0 \eta'
   + \del_1 \xi^I \eta_0 \eta
   + {\cal P}^I \eta_0 \eta'', 
\nonumber \\
s{\cal P}'^I 
&=& - \, P_{\bar{\phi}}^I 
    + \del_1 ({\cal P}^I \eta_0) 
    + \del_1 ({\cal P}'^I \eta_1), 
\nonumber \\
s\bar{\eta}
&=&- \, \del_1 (\eta_1 \bar{\eta})
   - \eta \del_1 \eta'
   - \eta' \del_1 \eta
   + \del_1 \eta'' , 
\nonumber \\
s\bar{{\cal P}}
&=& - \, \frac{1}{2} \phi^I \phi_I 
    + \phi^I {\cal P}'_I \eta_0 
    + \del_1 \bar{{\cal P}} \eta_1
    - {\cal P}'' \eta_0 \del_1 \eta_0 , 
\nonumber \\ 
s\eta''
&=&- \, \eta_1 \del_1 \eta''
   - \eta_0 \eta \del_1 \eta
   - \eta_0 \eta' \del_1 \eta'
   - \bar{\eta} \eta_0 \del_1 \eta_0 ,
\nonumber \\
s{\cal P}''
&=&  \del_1\bar{{\cal P}} 
   - \phi_I {\cal P}^I 
   - \del_1 ({\cal P}'' \eta_1)
   + {\cal P}^I {\cal P}'_I \eta_0. 
\nonumber  
\end{eqnarray}
By using the generalized Poisson brackets,   
a nilpotent BRST charge $\Omega_{\rm min}$, 
which realizes the BRST transformations 
$sX\equiv\{\Omega_{\rm min}, X\}$ 
for any canonical variables $X$, 
is defined by
\begin{eqnarray}
\Omega_{\rm min}=\int\! \dd x^1
&\bigg\{\,&
\eta_0
\bigg(\, \frac{1}{2}\Big(P^I_\xi P_{\xi I}
                    +\del_1\xi^I\del_1\xi_I
                \Big)
    +P^I_{\bar{\phi}}P_{\phi I} 
    +\del_1\bar{\phi}^I\del_1\bar{\phi}_I 
\bigg)
\nonumber \\ &\phantom{\bigg(\,}&
+\,\eta_1
\Big(P^I_\xi\del_1\xi_I 
    +P^I_{\bar{\phi}}\del_1\bar{\phi}_I 
    +P^I_{\phi}\del_1\phi_I 
\Big) 
\nonumber \\ &\phantom{\bigg(\,}& 
    +\,\eta\phi_I\del_1\xi^I 
    +\eta'\phi_IP^I_\xi 
    +\eta^I\del_1\phi_I 
    +\eta'_IP_{\bar{\phi}}^I 
    +\frac{1}{2}\bar{\eta}\phi^I\phi_I   
\nonumber \\ &\phantom{\bigg(\,}&
    +\eta''\Big(\del_1\bar{\cal P}-\phi_I{\cal P}^I\Big) 
\nonumber \\ &\phantom{\bigg(\,}&
    +P^I_\xi{\cal P}_I\eta_0\eta 
    -P^I_\xi{\cal P}'_I\eta_0\eta' 
    +\xi^I\del_1({\cal P}'_I\eta_0\eta)
    -\xi^I\del_1({\cal P}_I\eta_0\eta')
    -\phi^I{\cal P}'_I\eta_0\bar\eta 
\nonumber \\ &\phantom{\bigg(\,}&
+\,{\cal P}_0\Big(\eta_0\del_1\eta_1
          +\eta_1\del_1\eta_0
      \Big) 
+{\cal P}_1\Big(\eta_0\del_1\eta_0
          +\eta_1\del_1\eta_1
      \Big) 
\nonumber \\ &\phantom{\bigg(\,}&
+\,{\cal P}\Big(\eta_0\del_1\eta'  
        +\eta_1\del_1\eta
    \Big) 
+{\cal P}'\Big(\eta_0\del_1\eta
       +\eta_1\del_1\eta'
   \Big)
\nonumber \\ &\phantom{\bigg(\,}&
+\,{\cal P}_I\Big(\eta_0\del_1\eta'^I
          +\eta_1\del_1\eta^I
      \Big)
+{\cal P}'_I\Big(\eta_0\del_1\eta^I 
         +\eta_1\del_1\eta'^I
     \Big)
\nonumber \\ &\phantom{\bigg(\,}&
+\,\bar{{\cal P}}\Big(\eta'\del_1\eta
                +\eta\del_1\eta'
                +\del_1(\eta_1\bar\eta)
            \Big) 
+{\cal P}^I{\cal P}'_I\eta_0\eta''
\nonumber \\ &\phantom{\bigg(\,}&
+\,{\cal P}''\Big(\eta_1\del_1\eta''
          + \eta_0 \eta \del_1\eta
          + \eta_0 \eta' \del_1\eta'
          + \bar{\eta} \eta_0 \del_1\eta_0
      \Big)\bigg\}.
\end{eqnarray}

In order to fix the gauge, we extend the phase space further 
and introduce sets of canonical variables 
$\Big(\bar{\lambda}(x), \lambda(x)\Big)$ 
and $\Big(\bar{\rho}(x), \rho(x)\Big)$. 
Their statics are bosonic 
for $\Big(\bar{\lambda}(x), \lambda(x)\Big)$ and fermionic 
for $\Big(\bar{\rho}(x), \rho(x)\Big)$ and 
the canonical structures are defined by
\begin{equation}
\begin{array}{rcl}
\{\bar{\lambda}, \lambda\}&=&-\{\lambda, \bar{\lambda}\}=1, 
\\
\{\bar{\rho}, \rho\}&=&\{\rho, \bar{\rho}\}=-1.
\end{array}
\end{equation}
BRST transformations are also extended to these 
variables as 
\begin{equation}
\begin{array}{rclcrcl}
s\bar{\lambda}&=&\rho, &\qquad& s\rho&=&0, \\
s\bar{\rho}&=&\lambda, &\qquad& s\lambda&=&0. \\
\end{array}
\end{equation}
The corresponding extended BRST charge is given by 
\begin{equation}
\Omega = \Omega_{\rm min} + \Omega_{\rm nonmin}, 
\label{BRST-charge}
\end{equation}
$$
\Omega_{\rm nonmin} = - \int\! \dd x^1 \, \lambda \rho.
$$

Now the gauge-fixed action is obtained by a Legendre transformation 
from the Hamiltonian in the extended phase space, 
\begin{eqnarray}
S_{\rm gauge\hbox{-}fixed}
= \int\! \dd x^0 \bigg\{\int\! \dd x^1
&\Big(\,& \del_0\xi^I P_{\xi I}
       + \del_0\bar{\phi}^I P_{\bar{\phi}I} 
       + \del_0\phi^I P_{\phi I} 
\nonumber \\
&&+\, \del_0\eta_0 {\cal P}_0 + \del_0\eta_1 {\cal P}_1 
  + \del_0\eta {\cal P}
  + \del_0\eta' {\cal P}' 
\nonumber \\
&&+\, \del_0\eta^I {\cal P}_I + \del_0\eta'^I {\cal P}'_I 
  + \del_0\bar{\eta} \bar{{\cal P}} + \del_0\eta'' {\cal P}'' 
\nonumber \\
&&+\, \del_0\bar{\rho} \rho
  + \del_0\bar{\lambda}\lambda \Big)
- H_K \bigg\}, 
\end{eqnarray}
where $H_K$ is a gauge-fixed Hamiltonian expressed by using 
a gauge-fixing fermion $K$, 
\begin{equation} 
H_K = \{\Omega, K\}.  
\end{equation}
The gauge-fixed Hamiltonian $H_K$ consists of gauge-fixing terms 
and ghost parts only since the total Hamiltonian of the system has  
vanished. 
There is no systematic way to find $K$ so as to yield a covariant
expression. 
Here, however, we can use the result in the Lagrangian formulation 
as a clue.   
Actually we would like to show that the two formulations give an 
equivalent result. 
We have found that the following gauge-fixing fermion $K$ 
works as desired
\begin{equation}
K = \int\! \dd x^1 
           \Big(-{\cal P}_0 
                + \bar{\eta} \del_1\bar{\lambda}
                +\bar{\rho}{\cal P}'' 
           \Big).
\end{equation}
By integrating out the momentum variables 
$P_{\xi}^I(x)$, $P_{\phi}^I(x)$, $P_{\bar{\phi}}^I(x)$, 
${\cal P}''(x)$ and $\lambda(x)$ with this gauge-fixing fermion, 
we obtain the following relations, 
\begin{eqnarray}
P_{\xi}^I 
&=& \del_0\xi^I + {\cal P}^I\eta - {\cal P}'^I\eta' , 
\nonumber \\
P_{\bar{\phi}}^I
&=& \del_0\phi^I, 
\nonumber \\
P_{{\phi}}^I 
&=& \del_0\bar{\phi}^I, 
\\
\lambda
&=& \del_0\eta''-\eta_1\del_1\bar{\rho} 
   + \bar{\eta} \del_1\eta_0 
   - \eta \del_1\eta - \eta' \del_1\eta' ,
\nonumber \\
{\cal P}'' 
&=& \del_0\bar{\lambda}. 
\nonumber 
\end{eqnarray}
Then, the gauge-fixed action becomes
\begin{eqnarray}
S_{\rm gauge\hbox{-}fixed}
= \int\! \dd^2 x \bigg\{\,
&&\frac{1}{2} \del_0\xi^I \del_0\xi_I
 -\frac{1}{2} \del_1\xi^I \del_1\xi_I
\nonumber \\
&&+ \, \del_0\bar{\phi}^I \del_0\phi_I
  - \del_1\bar{\phi}^I \del_1\phi_I
\nonumber \\
&&- \, {\cal P}_0 \del_0\eta_0
  + {\cal P}_1 \del_1\eta_0 
  - {\cal P}_1 \del_0\eta_1
  + {\cal P}_0 \del_1\eta_1 
\nonumber \\
&&- \, {\cal P} \del_0\eta
  + {\cal P}' \del_1\eta 
  - {\cal P}' \del_0\eta'
  + {\cal P} \del_1\eta' 
\nonumber \\
&&- \, {\cal P}_I \del_0\eta^I
  + {\cal P}'_I \del_1\eta^I 
  - {\cal P}'_I \del_0\eta'^I
  + {\cal P}_I \del_1\eta'^I 
\nonumber \\
&&- \, \bar{{\cal P}} \del_0\bar{\eta}
  + \rho \del_1\bar{\eta}
  - \rho \del_0\bar{\rho}
  + \bar{\rho} \del_1\bar{{\cal P}} 
\nonumber \\
&&+ \, \del_0\bar{\lambda} \del_0\eta''
  - \del_1\bar{\lambda} \del_1\eta''
\nonumber \\
&& + \, \del_0\xi^I {\cal P}_I\eta 
   - \del_1\xi^I {\cal P}'_I\eta
   - \del_0\xi^I {\cal P}'_I\eta'
   + \del_1\xi^I {\cal P}_I\eta'
   - \phi^I {\cal P}'_I \bar{\eta}
   + \phi^I {\cal P}_I \bar{\rho} 
\nonumber \\
&&+ \, \del_1\bar{\lambda} \eta \del_1\eta'
  + \del_1\bar{\lambda} \eta' \del_1\eta 
  - \del_1\bar{\lambda} \del_1(\bar{\eta} \eta_1) 
\nonumber \\
&&- \, \del_0\bar{\lambda} \eta_1 \del_1\bar{\rho}
  + \del_0\bar{\lambda} \bar{\eta} \del_1\eta_0
  - \del_0\bar{\lambda} \eta \del_1\eta 
  - \del_0\bar{\lambda} \eta' \del_1\eta' 
\nonumber \\
&&- \, {\cal P}^I {\cal P}'_I \eta_0 \bar{\rho} 
  + {\cal P}^I {\cal P}'_I \eta \eta'
  - {\cal P}^I {\cal P}'_I \eta''
\bigg\}.
\label{gauge-fixed_03}
\end{eqnarray}
If we redefine the field variables as: 
$$
\begin{array}{rclcrcl}
\eta_0 &\rightarrow& 
-d^0, &\qquad
& {\cal P}_0 &\rightarrow&
-\bar{d}_{00} - \hat{c}^0 \del_1 \bar{f}
+ \hat{c}^1 \del_0 \bar{f}, 
\\
\eta_1 &\rightarrow&
-d^1, &\qquad
& {\cal P}_1 &\rightarrow&
-\bar{d}_{01} + \hat{c}^1 \del_1\bar{f}, 
\\
\eta &\rightarrow&
a, &\qquad
& {\cal P} &\rightarrow&
\hat{a}^0 - \del_1(\bar{f}a) + \bar{f} \del_0 a' + \hat{b}^0_I \xi^I, 
\\
\eta' &\rightarrow&
- a', &\qquad
& {\cal P}' &\rightarrow&
- \hat{a}^1 + \del_1(\bar{f} a') - \bar{f} \del_0 a - \hat{b}^1_I \xi^I, 
\\
\eta^I &\rightarrow&
b^I - a\xi^I, &\qquad
& {\cal P}^I &\rightarrow&
\hat{b}^{0I}, 
\\
\eta'^I &\rightarrow&
- b'^I + a' \xi^I, &\qquad
& {\cal P}'^I &\rightarrow&
- \hat{b}^{1I}, 
\\
\bar{\eta} &\rightarrow&
- \hat{c}^0, &\qquad
&\bar{{\cal P}} &\rightarrow&
- c + d^0 \del_1\bar{f}, 
\\
\eta'' &\rightarrow&
f + d^0 \hat{c}^1, &\qquad 
&\bar{\lambda} &\rightarrow&
\bar{f}, 
\\
\rho &\rightarrow&
c' + d^0 \del_0\bar{f} + d^1 \del_1 \bar{f}, &\qquad
&\bar{\rho}&\rightarrow&
\hat{c}^1, 
\end{array}
$$
the action (\ref{gauge-fixed_03}) and the BRST transformations 
(\ref{hamiltonian_BRST}) completely coincide with the gauge-fixed 
action (\ref{gauge-fixed_02}) and the on-shell BRST transformations 
(\ref{lagrangian_BRST}) in the Lagrangian formulation. 
After these manipulations we also obtain the final form of 
the BRST charge (\ref{BRST-charge}), 
\begin{eqnarray*}
\Omega 
= \int \!\dd x^1 
\bigg\{ 
&&- d^0 \Big(\, 
\frac{1}{2} \partial_0 \xi^I \partial_0 \xi_I
    + \frac{1}{2} \partial_1 \xi^I \partial_1 \xi_I
    + \partial_0 \bar{\phi}^I \partial_0 \phi_I
    + \partial_1 \bar{\phi}^I \partial_1 \phi_I
\nonumber\\
&& \hspace*{10mm}
    - \, \hat{a}^1 \partial_0 a'
    + \, \hat{a}^0 \partial_1 a'
    - \hat{b}^1_I \partial_0 b'^I
    + \hat{b}^0_I \partial_1 b'^I
    - \hat{c}^1 \partial_0 c'
    + \hat{c}^0 \partial_1 c'
\nonumber \\
&& \hspace*{10mm}
    + \partial_0 \bar{f} \partial_0 f
    + \partial_1 \bar{f} \partial_1 f 
        \Big)
\nonumber\\
&& - d^1 \Big( 
      \partial_0 \xi^I \partial_1 \xi_I
    + \partial_0 \bar{\phi}^I \partial_1 \phi_I
    + \partial_1 \bar{\phi}^I \partial_0 \phi_I
\nonumber\\
&& \hspace*{10mm}
    + \, \hat{a}^0 \partial_0 a'
    - \, \hat{a}^1 \partial_1 a'
    + \hat{b}^0_I \partial_0 b'^I
    - \hat{b}^1_I \partial_1 b'^I
    + \hat{c}^0 \partial_0 c'
    - \hat{c}^1 \partial_1 c'
\nonumber \\
&& \hspace*{10mm}
    + \partial_0 \bar{f} \partial_1 f
    + \partial_1 \bar{f} \partial_0 f 
          \Big)
\nonumber \\
&&           
    - \bar{d}_{0n} d^m \partial_m d^n
\nonumber \\
&&   - a' ( \phi_I \partial_0\xi^I - \partial_0 \phi_I \xi^I )
     + a ( \phi_I \partial_1 \xi^I - \partial_1\phi_I \xi^I )
\nonumber \\
&& 
     - b'_I \partial_0\phi^I
     + b_I  \partial_1\phi^I
     - \frac{1}{2} \hat{c}^0 \phi^I\phi_I
\nonumber \\
&& 
     - c' ( \partial_0 f + a \partial_0 a' + a' \partial_0 a )
     + c  ( \partial_1 f + a \partial_1 a' + a' \partial_1 a )
     - (f+aa') \hat{b}^0_I \phi^I
\bigg\}. 
\end{eqnarray*}
%

%%%%%
\section{Noncovariant quantization in the light-cone gauge formulation}
\setcounter{equation}{0}
\setcounter{footnote}{0}

In this section 
we investigate the dynamics of the model 
defined by the constraints (\ref{primary}) and 
(\ref{1st_class_1})-(\ref{1st_class_7}) and 
Hamiltonian (\ref{hamiltonian}) in 
noncovariant gauge and 
obtain the same result of the critical dimension as 
in the covariant quantization\footnote{
%%%%% footnote %%%%% 
In this section, 
we use conventions of the world-sheet coordinates as 
$x^0\equiv\tau$ and $x^1\equiv\sigma$. 
We also parameterize the spatial coordinate 
as $0\le\sigma\le 2\pi$ and impose the periodical 
boundary conditions on any fields 
$\Phi^A(\tau, \sigma)$  
as $\Phi^A(\tau, \sigma)=\Phi^A(\tau, \sigma+2\pi)$. 
%%%%%
}. 
In addition, we present a mass-shell relation of the model 
and give low energy quantum states.    
According to imposing the noncovariant gauge fixing conditions, 
we explicitly solve the constraints to some of the variables 
from the equations of motion. 

We begin by considering conditions for the scalar field 
$\phi^I(\tau, \sigma)$. 
It is convenient to introduce Fourier mode expansions  
of the canonical pair 
$\Big(\phi^I(\tau, \sigma), P_\phi^J(\tau, \sigma)\Big)$, 
\begin{equation}
\begin{array}{rcl}
\phi^I(\tau, \sigma)
&=& \displaystyle 
    \phi^I(\tau)+\frac{1}{\sqrt{2\pi}} \sum_{m \ne 0}
                   \phi^I_m(\tau) e^{im\sigma}, \\
P_\phi^I(\tau, \sigma)
&=& \displaystyle
   \frac{p^I_\phi(\tau)}{2\pi} 
   + \frac{1}{\sqrt{2\pi}} \sum_{m \ne 0}
     p_{\phi m}^I(\tau) e^{im\sigma}.
\end{array}
\end{equation}
Poisson brackets are defined by 
\begin{eqnarray}
\{\phi^I(\tau), p^J_\phi(\tau)\} 
&=& \eta^{IJ}, \nonumber \\
\{\phi_m^I(\tau), p_{\phi n}^J(\tau)\} 
&=& \eta^{IJ}\delta_{m+n}, \\
\hbox{otherwise} &=& 0. \nonumber
\end{eqnarray}
In terms of the Fourier modes, 
the constraint (\ref{1st_class_5}) is 
equivalent to $\phi^I_m(\tau)=0$.  
We will later adopt a gauge fixing condition for this constraint. 
On the other hand, 
the equation of motion for $\phi^I(\tau, \sigma)$ 
on the constraint surface is 
$\del_\tau\phi^I(\tau, \sigma) = 0$. 
Together with the constraint $\phi_m^I(\tau)=0$, 
we then set the configuration of the scalar field 
as $\phi^I(\tau, \sigma)=\phi^I(\tau)=\phi^I(=\mbox{const.})$.

As we did in the previous section, 
by using the gauge parameters $k^n(\tau, \sigma)$ 
for the general coordinate transformations
we first adopt gauge fixing conditions for the constraints 
$P_N(\tau, \sigma)=0$ and $P_{N_1}(\tau, \sigma)=0$ as 
the orthonormal gauge $N(\tau, \sigma)=1$ 
and $N_1(\tau, \sigma)=0$. 
The {\UvUa} gauge parameters $v(\tau, \sigma)$, $v'(\tau, \sigma)$ 
and the global parameters $\alpha_i$ 
can fix to be $A_m(\tau, \sigma)=0$ corresponding 
to the constraints $P_{A_m}(\tau, \sigma)=0$. 
However, the system still has residual symmetries 
concerned with these gauge parameters $k^n(\tau, \sigma)$,  
$v(\tau, \sigma)$ and $v'(\tau, \sigma)$. 
Taking these symmetries into account, 
we can adopt the following gauge fixing conditions 
on ``two'' light-cone coordinates\footnote{
%%%%% footnote %%%%%
From the definition of the metric (\ref{FlatmetricDefinitionGravity}), 
we denote the light-cone coordinates of the 
background spacetime as 
$x^I =(x^+, x^-, x^i, x^{\hat{+}}, x^{\hat{-}})$,   
where $x^{\pm}\equiv\frac{1}{\sqrt{2}}(x^0\pm x^{D-3})$ and 
$x^{\hat{\pm}}\equiv\frac{1}{\sqrt{2}}(x^{\hat{0}}\pm x^{\hat{1}})$ and  
the index $i$ runs through $1$, $2$, \ldots, $D-4$. 
} 
%%%%%
of the background spacetime 
within the gauge $N(\tau, \sigma)=1$, $N_1(\tau, \sigma)=0$ 
and $A_m(\tau, \sigma)=0$, 
%
%%% new arraystretch
\renewcommand{\arraystretch}{2.2}
%%%
%
\begin{equation}
\begin{array}{rclcrcl}
\xi^+ (\tau, \sigma)
&=& \displaystyle\frac{p^+}{2\pi} \tau, &\qquad&
P_\xi^+ (\tau, \sigma)
&=& \displaystyle\frac{p^+}{2\pi}, \\
\xi^{\hat{+}} (\tau, \sigma)
&=& \displaystyle\frac{p^{\hat{+}}}{2\pi} \tau, &\qquad&
P_\xi^{\hat{+}} (\tau, \sigma)
&=& \displaystyle\frac{p^{\hat{+}}}{2\pi}, 
\end{array}
\label{light-cone_gauge_fixing_1}
\end{equation}
%
%%% new arraystretch
\renewcommand{\arraystretch}{1.6}
%%%

\vspace*{-3mm}
\noindent
where $p^+$ and $p^{\hat{+}}$ are light-cone components of 
the center of mass momenta. 
Therefore we can eliminate ``two'' unphysical components of  
the coordinates of the background spacetime. 
Indeed the gauge fixing 
conditions (\ref{light-cone_gauge_fixing_1}) correspond to ones for 
the first-class constraints (\ref{1st_class_1})-(\ref{1st_class_4}). 

In order to show how these 
conditions (\ref{light-cone_gauge_fixing_1}) are accomplished, 
we use Fourier mode expansions of the canonical pair 
$\Big(\xi^I(\tau, \sigma), P_\xi^J(\tau, \sigma)\Big)$.  
Under the gauge $N(\tau, \sigma)=1$, $N_1(\tau, \sigma)=0$ 
and $A_m(\tau, \sigma)=0$, 
the equations of motion for $\xi^I(\tau, \sigma)$ 
and $P_\xi^I(\tau, \sigma)$ 
turn to be free wave equations and their solutions are 
\begin{equation}
\begin{array}{rcl}
\xi^I(\tau, \sigma) 
&=& \displaystyle 
    x^I + \frac{p^I}{2\pi} \tau 
    +\frac{i}{2\sqrt{\pi}} \sum_{m \ne 0}
     \frac{1}{m}\Big(\alpha^I_m e^{-im(\tau-\sigma)}
                     + \tilde{\alpha}^I_m e^{-im(\tau+\sigma)}
                \Big), \\
P_\xi^I(\tau, \sigma)
&=& \displaystyle 
    \frac{p^I}{2\pi}
    +\frac{1}{2\sqrt{\pi}} \sum_{m \ne 0}
    \Big(\alpha^I_me^{-im(\tau-\sigma)}
         + \tilde{\alpha}^I_m e^{-im(\tau+\sigma)}
    \Big), 
\end{array}
\end{equation}
and Poisson brackets are given by 
\begin{eqnarray}
\{x^I, p^J\} 
&=& \eta^{IJ}, \nonumber \\
\{\alpha^I_m, \alpha^J_n\} 
&=& \{\tilde{\alpha}^I_m, \tilde{\alpha}^J_n\}
   = - i m \eta^{IJ} \delta_{m+n}, \\
\hbox{otherwise} 
&=& 0. \nonumber
\end{eqnarray}
In terms of the Fourier modes, 
the constraints (\ref{1st_class_1})-(\ref{1st_class_4}) are 
equivalent to 
\begin{subequations}
\begin{eqnarray}
&&L_m = \tilde{L}_m = 0, 
\label{virasoro} \\
&&\phi_I \alpha_m^I = \phi_I \tilde{\alpha}_m^I = 0, 
\label{additional_constraints}
\end{eqnarray}
\end{subequations}

\vspace*{-8mm}
\noindent
where we define the Virasoro generators as 
$$
L_m
\equiv \frac{1}{2}\sum_k\alpha_{m-k}^I{\alpha_I}_k, \qquad
\tilde{L}_m
\equiv \frac{1}{2}\sum_k{\tilde{\alpha}}_{m-k}^I\tilde{\alpha}_{Ik}, 
$$
and we denote 
$\alpha_0^I=\tilde{\alpha}^I_0\equiv p^I/(2\sqrt{\pi})$.
The gauge fixing conditions 
(\ref{light-cone_gauge_fixing_1}) are equivalent to 
\begin{equation}
\begin{array}{l} 
x^+ = x^{\hat{+}} = 0,  
\\
\alpha^{+}_m = \alpha^{\hat{+}}_m =
\tilde{\alpha}^{+}_m = \tilde{\alpha}^{\hat{+}}_m = 0,  
\qquad (m\ne 0).
\end{array}
\label{light-cone_gauge_fixing_fourier_1}
\end{equation}

Now let us explain the procedure to obtain the gauge fixing 
conditions (\ref{light-cone_gauge_fixing_fourier_1}).  
Within the orthonormal gauge 
we can perform changes of the background spacetime coordinates 
with the gauge parameters $k^n(\tau, \sigma)$ provided that conditions 
$\del_\tau k^\tau(\tau, \sigma)=\del_\sigma k^\sigma(\tau, \sigma)$ 
and $\del_\tau k^\sigma(\tau, \sigma)=\del_\sigma k^\tau(\tau, \sigma)$ 
are satisfied. 
Here we take the following parameterizations of $k^n(\tau, \sigma)$ 
which satisfy these conditions, 
\begin{eqnarray*}
k^+(\tau, \sigma) 
&\equiv& \displaystyle \frac{1}{\sqrt{2}}(k^\tau + k^\sigma) 
= \displaystyle \frac{1}{\sqrt{2}}\sum_m \tilde{k}_m
                  e^{-im(\tau+\sigma)}, \\
k^-(\tau, \sigma) 
&\equiv& \displaystyle \frac{1}{\sqrt{2}}(k^\tau - k^\sigma) 
= \displaystyle \frac{1}{\sqrt{2}} \sum_m k_m 
                  e^{-im(\tau-\sigma)}. 
\end{eqnarray*}
In addition to these, the {\UvUa} gauge 
parameters $v(\tau, \sigma)$ and $v'(\tau, \sigma)$ can 
be also used to perform changes of the 
coordinates within the gauge $A_m(\tau, \sigma)=0$  
provided that conditions 
$\del_\tau v'(\tau, \sigma)=-\del_\sigma v(\tau, \sigma)$ 
and $\del_\tau v(\tau, \sigma) =-\del_\sigma v'(\tau, \sigma)$ 
are satisfied. 
We take the following parameterizations of $v(\tau, \sigma)$ 
and $v'(\tau, \sigma)$ 
to realize these conditions, 
\begin{eqnarray*}
v(\tau, \sigma) 
&=& v + \displaystyle \frac{i}{2\sqrt{\pi}} \sum_{m\ne 0}
                      \frac{1}{m}\Big(v_m e^{-im(\tau-\sigma)}
                                     -\tilde{v}_m e^{-im(\tau+\sigma)}
                                 \Big), \\
v'(\tau, \sigma)
&=& v' + \displaystyle \frac{i}{2\sqrt{\pi}} \sum_{m \ne 0}
                      \frac{1}{m}\Big(v_m e^{-im(\tau-\sigma)}
                                     +\tilde{v}_me^{-im(\tau+\sigma)}
                                 \Big). 
\end{eqnarray*}
The gauge transformations corresponding to these parameters 
are consistent with the equations of motion for $\xi^I(\tau, \sigma)$ 
and $P_\xi^I(\tau, \sigma)$. 
Because, in terms of the Fourier modes, 
the gauge transformations are given by 
\begin{equation}
\begin{array}{rcl}
\delta x^I 
&=& \displaystyle 
   \frac{1}{2\sqrt{\pi}} \sum_m k_m \alpha_{-m}^I 
   + \frac{1}{2\sqrt{\pi}} \sum_m \tilde{k}_m \tilde{\alpha}_{-m}^I 
   + v'\phi^I, 
\\
\delta p^I 
&=& 0, 
\\
\delta \alpha^I_m 
&=& \displaystyle 
    - i m \sum_n k_{m-n} \alpha_n^I
    + v_m \phi^I, \qquad (m\ne 0), 
\\
\delta \tilde{\alpha}^I_m 
&=& \displaystyle 
    - i m \sum_n \tilde{k}_{m-n} \tilde{\alpha}_n^I
    + \tilde{v}_m \phi^I, \qquad (m\ne 0).
\end{array}
\label{gauge_trans_fourier_1}
\end{equation}
It is worth to mention that these gauge transformations  
are the same ones 
in usual string theories, 
except for the gauge transformations corresponding to the parameters 
$v'$, $v_m$ and $\tilde{v}_m$. 
However, we would like to emphasize that these gauge transformations  
can be disappear on the following components, 
\begin{eqnarray*}
\delta (\phi^{\hat{+}}x^+-\phi^+x^{\hat{+}})
&=&\frac{1}{2\sqrt{\pi}}
    \sum_{m}\Big(
    \phi^{\hat{+}}(k_m\alpha_{-m}^+
                  +\tilde{k}_m\tilde{\alpha}_{-m}^+
                  )
   -\phi^+(k_m\alpha_{-m}^{\hat{+}}
          +\tilde{k}_m\tilde{\alpha}_{-m}^{\hat{+}}
          )
            \Big), 
\\
\delta (\phi^{\hat{+}}\alpha^{+}_m -\phi^+\alpha^{\hat{+}}_m) 
&=& -im\sum_{n}k_{m-n}(\phi^{\hat{+}}\alpha^+_n 
                      -\phi^+\alpha^{\hat{+}}_n
                      ), \qquad (m\ne 0),  
\\
\delta (\phi^{\hat{+}}\tilde{\alpha}^{+}_m 
       -\phi^+\tilde{\alpha}^{\hat{+}}_m) 
&=& -im\sum_{n}\tilde{k}_{m-n}(\phi^{\hat{+}}\tilde{\alpha}^+_n 
                              -\phi^+\tilde{\alpha}^{\hat{+}}_n
                               ), \qquad (m\ne 0).
\end{eqnarray*}
By using the gauge degrees of freedom for $k_m$ and $\tilde{k}_m$, 
which is the same manipulation to realize the light-cone gauge fixing 
condition in usual string theories,  
we can adopt gauge conditions  
\begin{eqnarray}
\phi^{\hat{+}}x^+-\phi^+x^{\hat{+}} &=& 0, 
\nonumber \\
\phi^{\hat{+}}\alpha_m^+-\phi^+\alpha_m^{\hat{+}} 
&=& 0, \qquad (m\ne 0),  
\label{light-cone_gauge_fixing_fourier_1-1} \\
\phi^{\hat{+}}\tilde{\alpha}_m^+-\phi^+\tilde{\alpha}_m^{\hat{+}} 
&=& 0, \qquad (m\ne 0),  
\nonumber 
\end{eqnarray}
if the following condition is satisfied, 
\begin{equation}
\phi^{\hat{+}}p^+-\phi^+p^{\hat{+}} \ne 0.
\label{light-cone_ass}
\end{equation}
Next we use the gauge degrees of freedom for $v'$, $v_m$ and 
$\tilde{v}_m$ in (\ref{gauge_trans_fourier_1}). 
To keep the condition (\ref{light-cone_ass}) 
both of the scalar fields $\phi^{\hat{+}}$ and $\phi^+$ can 
not be vanish simultaneously. 
If $\phi^{\hat{+}}\ne 0$, 
we can adopt the following gauge fixing conditions 
of the $\hat{+}$ component,  
\begin{equation}
\begin{array}{l}
x^{\hat{+}} = 0, 
\\
\alpha^{\hat{+}}_m=\tilde{\alpha}^{\hat{+}}_m=0, \qquad (m\ne 0), 
\end{array}
\label{light-cone_gauge_fixing_fourier_1-2}
\end{equation}
without spoiling the gauge fixing 
conditions (\ref{light-cone_gauge_fixing_fourier_1-1}).  
From (\ref{light-cone_gauge_fixing_fourier_1-1}) and 
(\ref{light-cone_gauge_fixing_fourier_1-2}) we can then obtain 
the gauge fixing conditions (\ref{light-cone_gauge_fixing_fourier_1}). 
In the similar way, we also conclude the same gauge fixing conditions 
(\ref{light-cone_gauge_fixing_fourier_1}), in the case $\phi^+\ne 0$. 
Therefore without the loss of the generality 
we choose the case $\phi^{\hat{+}} \ne 0$ 
throughout the rest of this paper. 

We next adopt the gauge fixing 
condition 
$C_{\tau\sigma}(\tau, \sigma)=-\hat{C}(\tau, \sigma)=-\hat{C}_0$ \ (const.) 
with respect to constraints $P_{C_{\tau\sigma}}(\tau, \sigma)=0$, 
by using the gauge parameter $w_m(\tau, \sigma)$.    
Within this gauge we also have a residual gauge symmetry 
corresponding to the gauge parameter $w_m(={\hbox{const.}})$ 
which will be used later. 

Using the remaining gauge parameters $u^I(\tau, \sigma)$ 
and $u'^I(\tau, \sigma)$, 
we can further impose gauge fixing conditions for the constraints 
$P_{\bar{\phi}}^I(\tau, \sigma)=0$, 
$P_{B_m}^I(\tau, \sigma)=0$ and 
$\del_\sigma\phi^I(\tau, \sigma)=0$. 
In order to specify the gauge fixing condition 
it is also convenient to introduce Fourier mode expansions. 
We list below these for the canonical pairs 
$\Big(\bar{\phi}^I(\tau, \sigma), P_{\bar{\phi}}^J(\tau, \sigma)\Big)$ 
and $\Big(B_m^I(\tau, \sigma), P_{B_m}^J(\tau, \sigma)\Big)$ 
and their Poisson brackets:
\begin{itemize}
\item $(\bar{\phi}^I, P_{\bar{\phi}}^J)$ sector: 
\begin{equation}
\begin{array}{rcl}
\bar{\phi}^I(\tau, \sigma)
&=& \displaystyle 
    \bar{\phi}^I(\tau)
   +\frac{1}{\sqrt{2\pi}}\sum_{m\ne 0}
    \bar{\phi}^I_m(\tau)e^{im\sigma}, \\
P_{\bar{\phi}}^I(\tau, \sigma)
&=& \displaystyle 
   \frac{p_{\bar{\phi}}^I(\tau)}{2\pi}
   +\frac{1}{\sqrt{2\pi}}\sum_{m\ne 0}
    p_{\bar{\phi} m}^I(\tau)e^{im\sigma}, 
\end{array}
\end{equation}
and Poisson brackets,  
\begin{eqnarray}
\{\bar{\phi}^I(\tau), p^J_{\bar{\phi}}(\tau)\} 
&=& \eta^{IJ}, \nonumber \\
\{\bar{\phi}_m^I(\tau), p_{\bar{\phi} n}^J(\tau)\} 
&=& \eta^{IJ}\delta_{m+n}, \\
\hbox{otherwise} &=& 0. \nonumber 
\end{eqnarray}
\item $(B_\tau^I, P_{B_\tau}^J)$ sector: 
\begin{equation}
\begin{array}{rcl}
B_\tau^I(\tau, \sigma) 
&=& \displaystyle
    {B_{\tau}^I(\tau)} 
   +\frac{1}{\sqrt{2\pi}}\sum_{m\ne 0}
    B_{\tau m}^I(\tau)e^{im\sigma}, \\
P_{B_\tau}^I(\tau, \sigma) 
&=& \displaystyle
    \frac{p_{B_\tau}^I(\tau)}{2\pi}
   +\frac{1}{\sqrt{2\pi}}\sum_{m\ne 0}
    p_{B_\tau m}^I(\tau)e^{im\sigma}, 
\end{array}
\end{equation}
and Poisson brackets,  
\begin{eqnarray}
\{B_{\tau}^I(\tau), p_{B_\tau}^J(\tau)\} 
&=&\eta^{IJ}, \nonumber \\
\{B_{\tau m}^I(\tau), p_{B_\tau n}^J(\tau)\}
&=&\eta^{IJ}\delta_{m+n}, \\
\hbox{otherwise} 
&=&0, \nonumber
\end{eqnarray}
and the similar relations for $(B_\sigma^I, P_{B_\sigma}^J)$. 
\end{itemize}
Then, equations of motion for 
$\bar{\phi}^I(\tau, \sigma)$ and 
$P_\phi^I(\tau, \sigma)$, 
\begin{equation}
\begin{array}{rcl}
\del_\tau\bar{\phi}^I(\tau, \sigma) 
&=& P_\phi^I(\tau, \sigma)+B_\sigma^I(\tau, \sigma), \\
\del_\tau P_\phi^I(\tau, \sigma) 
&=& \del^2_\sigma\bar{\phi}^I(\tau, \sigma) 
   -\del_\sigma B_\tau^I(\tau, \sigma) 
   -\hat{C}_0\phi^I, 
\end{array}
\end{equation}
are expressed by the Fourier modes,  
\begin{subequations}
\begin{eqnarray}
\del_\tau\bar{\phi}^I (\tau)
&=& \frac{p_\phi^I(\tau)}{2\pi} 
   +B^I_{\sigma}(\tau),
\label{eq_motion_1} 
\\
\del_\tau\bar{\phi}^I_m (\tau) 
&=& p^I_{\phi m}(\tau) + B^I_{\sigma m}(\tau), 
\label{eq_motion_2}
\\
\del_\tau p_{\phi}^I (\tau) 
&=& - 2 \pi \hat{C}_0\phi^I, 
\label{eq_motion_3}
\\
\del_\tau p_{\phi m}^I (\tau)
&=& -m^2\bar{\phi}^I_m(\tau) 
    -imB_{\tau m}^I(\tau). 
\label{eq_motion_4}
\end{eqnarray}
\end{subequations}

\vspace*{-8mm}
\noindent
The equation of motion (\ref{eq_motion_3}) for the non-oscillator mode 
of $P_\phi^I(\tau, \sigma)$ can be solved as 
\begin{equation}
p_\phi^I(\tau) = p_\phi^I - 2\pi \hat{C}_0 \phi^I \tau,
\end{equation}
where $p_\phi^I$ is a zero-mode and Poisson bracket is defined 
by
\begin{equation}
\{\phi^I, p_\phi^J\}=\eta^{IJ}.
\end{equation}
On the Fourier components, the constraints 
$P_{\bar{\phi}}^I(\tau, \sigma)=0$, $P_{B_m}^I(\tau, \sigma)=0$ and 
$\del_\sigma\phi^I(\tau, \sigma)=0$ are equivalent 
to 
\begin{equation}
\begin{array}{rcl}
p_{\bar{\phi}}^I(\tau)=p_{\bar{\phi} m}^I(\tau)&=& 0, \\
p_{B_\tau}^I(\tau)=p_{B_{\tau} m}^I(\tau)&=&0, \\
p_{B_\sigma}^I(\tau)=p_{B_{\sigma} m}^I(\tau)&=&0, \\
\phi_m^I(\tau)&=&0. 
\end{array}
\label{1st_class_fourier_2}
\end{equation}
Now we impose gauge fixing conditions 
corresponding to the constraints (\ref{1st_class_fourier_2}). 
The gauge fixing conditions 
are determined so as to be compatible with the equations of motion 
(\ref{eq_motion_1})-(\ref{eq_motion_4}). 
By making the gauge transformations 
\begin{eqnarray*}
\delta\bar{\phi}{}^I 
&=& u'^I, 
\\
\delta B_\tau^I 
&=& \del_\tau u^I + \del_\sigma u'^I, 
\\
\delta B_\sigma^I 
&=& \del_\tau u'^I + \del_\sigma u^I, 
\\
\delta P_\phi^I 
&=& -\del_\sigma u^I, 
\end{eqnarray*}
with gauge parameters  
\begin{eqnarray*}
u^I (\tau, \sigma) 
&=& -\int^\tau \!\! \dd\tau' B_{\tau}^I(\tau')
- \frac{i}{\sqrt{2\pi}}\sum_{m\ne 0}\frac{1}{m}
    p_{\phi m}^I(\tau)e^{im\sigma}, \\
u'^I (\tau, \sigma)
&=& - \,\bar{\phi}^I(\tau) 
       -\frac{1}{\sqrt{2\pi}}\sum_{m\ne 0}\bar{\phi}_m^I(\tau) 
        e^{im\sigma},
\end{eqnarray*}
and the equations of motion (\ref{eq_motion_1})-(\ref{eq_motion_4}), 
we obtain the following gauge fixing conditions, 
\begin{equation}
\begin{array}{rcl}
\bar{\phi}^I(\tau)= \bar{\phi}^I_m(\tau) &=& 0, \\
B_\tau^I (\tau) = B_{\tau m}^I(\tau) &=& 0, \\
B_\sigma^I(\tau) = \displaystyle - \,\frac{p_\phi^I(\tau)}{2\pi}, \qquad 
B_{\sigma m}^I(\tau) &=&0, \\
p_{\phi m}^I(\tau) &=& 0.  
\end{array}
\label{light-cone_gauge_fixing_fourier_2}
\end{equation}

Finally we consider the constraint 
\begin{equation}
\frac{1}{2}\phi^I\phi_I=0.
\label{1st_class_fourier_3}
\end{equation}
As we explained,  
the model has still residual gauge 
symmetry $w_\sigma(={\hbox{const.}})$ 
within the gauge $C_{\tau\sigma}(\tau, \sigma)=-\hat{C}_0$. 
Using this symmetry 
$$
\delta P_\phi^I=\phi^Iw_\sigma, 
$$
we can make the one of the zero-mode 
components of $P_\phi^I(\tau, \sigma)$ to be vanish. 
By taking the case $\phi^{\hat{+}}\ne 0$ 
and choosing the gauge parameter $w_\sigma$ as 
$$
w_\sigma = - \, \frac{p^{\hat{+}}_\phi}{2\pi\phi^{\hat{+}}}, 
$$  
we impose a gauge fixing condition 
\begin{equation}
p_\phi^{\hat{+}}=0.
\label{light-cone_gauge_fixing_fourier_3}
\end{equation}

We shall here summarize the correspondence between 
the constraints (\ref{virasoro}), (\ref{additional_constraints}),   
(\ref{1st_class_fourier_2}) and 
(\ref{1st_class_fourier_3}) and the gauge fixing conditions 
(\ref{light-cone_gauge_fixing_fourier_1}), 
(\ref{light-cone_gauge_fixing_fourier_2}) and 
(\ref{light-cone_gauge_fixing_fourier_3}) obtained from 
the above manipulation within the gauge 
$N(\tau, \sigma)=1$, $N_1(\tau, \sigma)=0$, 
$A_m(\tau, \sigma)=0$ and $C_{\tau\sigma}(\tau, \sigma)=-\hat{C}_0$:     
\begin{eqnarray*}
{\hbox{\bf constraints}} &\qquad\qquad& {\hbox{\bf gauge fixing conditions}} 
\nonumber \\
L_0 + \tilde{L}_0 = 0, 
&\qquad\qquad& x^+ = 0, 
\nonumber \\
L_m =\tilde{L}_m = 0,
&\qquad\qquad& 
\alpha^{+}_m=\tilde{\alpha}^{+}_m=0, \quad (m\ne 0), 
\nonumber \\
\phi_I p^I = 0, 
&\qquad\qquad& x^{\hat{+}} = 0, 
\nonumber \\
\phi_I \alpha_m^I = \phi_I {\tilde{\alpha}}_m^I = 0,  
&\qquad\qquad& \alpha^{\hat{+}}_m={\tilde{\alpha}}^{\hat{+}}_m = 0, 
\quad (m\ne 0), 
\nonumber \\ 
p_{\bar{\phi}}^I(\tau)=p_{\bar{\phi} m}^I(\tau)=0, 
&\qquad\qquad& \bar{\phi}^I(\tau)=\bar{\phi}_m^I(\tau)=0, 
\\ 
p_{B_\tau}^I(\tau)=p_{B_\tau m}^I(\tau)=0, 
&\qquad\qquad&
B_\tau^I(\tau)=B_{\tau m}^I(\tau)=0, 
\nonumber \\
p_{B_\sigma}^I(\tau)=p_{B_\sigma m}^I(\tau)=0, 
&\qquad\qquad& B_{\sigma}^I(\tau)= - \,\frac{p_\phi^I(\tau)}{2\pi}, 
\qquad B_{\sigma m}^I(\tau)=0,   
\nonumber \\
\phi_m^I(\tau)=0, 
&\qquad\qquad& p_{\phi m}^I(\tau)=0, 
\nonumber \\
\frac{1}{2}\phi^I\phi_I = 0, 
&\qquad\qquad& p^{\hat{+}}_\phi = 0. 
\nonumber
\end{eqnarray*}
Under these gauge fixing conditions, 
the dynamics of the model is described by 
the zero-modes and the oscillator modes of the transverse 
string coordinates $\xi^i(\tau, \sigma)$, 
the zero-modes of light-cone coordinates $\xi^{\hat{\pm}}(\tau, \sigma)$ 
and $\xi^{\pm}(\tau, \sigma)$ 
and the zero-modes of the fields $\phi^I(\tau, \sigma)$ and 
$P_\phi^I(\tau, \sigma)\Big(=-B_\sigma^I(\tau, \sigma)\Big)$.    

In fact these gauge conditions completely fix the gauge degrees of 
freedom and these are consistent with the equations of motion. 
As the constraints are quadratic in the Fourier modes, we can solve 
the constraints directly and the dependent variables are expressed 
in terms of the independent variables. 
Here are the independent canonical variables 
\begin{eqnarray}
\{x^-, p^+\} &=& \{x^{\hat{-}}, p^{\hat{+}}\} = -1, 
\nonumber \\
\{x^i, p^j\} &=& \delta^{ij}, 
\nonumber \\
\{\alpha^i_m, \alpha^j_n\} 
&=& \{\tilde{\alpha}^i_m, \tilde{\alpha}^j_n\} 
= -i m \delta^{ij} \delta_{m+n},  
\label{independent_variables} \\
\{\phi^+, p_\phi^-\} 
&=& 
\{\phi^-, p_\phi^+\} = \{\phi^{\hat{+}}, p_\phi^{\hat{-}}\} = -1, 
\nonumber \\
\{\phi^i, p_\phi^j\} &=& \delta^{ij}, 
\nonumber 
\end{eqnarray}
and the remaining non-vanishing dependent variables are 
\begin{eqnarray}
p^-
&=& \frac{-1}{\phi^{\hat{+}}p^+-\phi^+p^{\hat{+}}}
    \bigg\{\frac{p^{\hat{+}}p^{\hat{+}}}{\phi^{\hat{+}}}
          \Big(\phi^+\phi^--\frac{1}{2}\phi^i\phi_i
          \Big) 
\nonumber \\
&& \hspace*{30mm}
          -p^{\hat{+}}
          \Big(\phi^-p^+-\phi^ip_i
          \Big)
          -2\pi\phi^{\hat{+}}
          \Big(L_0^{\rm tr}+\tilde{L}_0^{\rm tr}
          \Big) 
    \bigg\}, 
\nonumber \\
\alpha^-_m 
&=& \frac{-1}{\phi^{\hat{+}}p^+-\phi^+p^{\hat{+}}}
    \Big(p^{\hat{+}}\phi^i\alpha_{m i} 
         -2\sqrt{\pi}\phi^{\hat{+}}L_{m}^{\rm tr}
    \Big), \qquad (m \ne 0), 
\nonumber \\
\tilde{\alpha}^-_m 
&=& \frac{-1}{\phi^{\hat{+}}p^+-\phi^+p^{\hat{+}}}
    \Big(p^{\hat{+}}\phi^i\tilde{\alpha}_{m i} 
         -2\sqrt{\pi}\phi^{\hat{+}}\tilde{L}_{m}^{\rm tr}
    \Big), \qquad (m \ne 0), 
\nonumber \\
p^{\hat{-}} 
&=& \frac{1}{\phi^{\hat{+}}p^+-\phi^+p^{\hat{+}}}
    \bigg\{\frac{p^{\hat{+}}p^+}{\phi^{\hat{+}}}
         \Big(\phi^+\phi^--\frac{1}{2}\phi^i\phi_i
         \Big) 
\label{dependent_variables} \\
&& \hspace*{30mm} 
    -p^+\Big(\phi^-p^+-\phi^ip_i
         \Big) 
     -2\pi\phi^+\Big(L_0^{\rm tr}+\tilde{L}_0^{\rm tr}
                \Big)
    \bigg\}, 
\nonumber \\
\alpha^{\hat{-}}_m 
&=& \frac{1}{\phi^{\hat{+}}p^+-\phi^+p^{\hat{+}}}
    \Big(p^+\phi^i\alpha_{m i} 
         -2\sqrt{\pi}\phi^+L_{m}^{\rm tr}
    \Big), \qquad (m\ne 0), 
\nonumber \\
\tilde{\alpha}^{\hat{-}}_m 
&=& \frac{1}{\phi^{\hat{+}}p^+-\phi^+p^{\hat{+}}}
    \Big(p^+\phi^i\tilde{\alpha}_{m i} 
         -2\sqrt{\pi}\phi^+\tilde{L}_{m}^{\rm tr}
    \Big), \qquad (m\ne 0), 
\nonumber \\
\phi^{\hat{-}} 
&=& - \, \frac{1}{\phi^{\hat{+}}}
    \Big(\phi^+\phi^--\frac{1}{2}\phi^i\phi_i
    \Big), 
\nonumber 
\end{eqnarray}
where the transverse parts of the Virasoro generators 
$L_m^{\rm tr}$ and $\tilde{L}_m^{\rm tr}$  are 
defined by
$$
L_m^{\rm tr}
\equiv\frac{1}{2}
\sum_k\alpha^i_{m-k}\alpha_{ik},
\qquad 
\tilde{L}_m^{\rm tr}
\equiv\frac{1}{2}
\sum_k\tilde{\alpha}^i_{m-k}\tilde{\alpha}_{ik}. 
$$

Now let us investigate the symmetry of the $D$-dimensional 
background spacetime. 
The translation and the Lorentz transformation generators 
derived from the classical action 
(\ref{classical_action_02}) are given by 
\begin{subequations}
\begin{eqnarray}
P^I &\equiv& \int_0^{2\pi} \!\!\! \dd \sigma \, P_\xi^I  
\nonumber \\
&=& p^I, 
\\
M^{IJ}&\equiv&\int_0^{2\pi} \!\!\! \dd \sigma \, 
              \Big(\xi^IP_\xi^J
                  +\phi^IP_\phi^J
                  +\bar{\phi}^IP_{\bar{\phi}}^J  
                  +B_\tau^IP_{B_\tau}^J 
                  +B_\sigma^IP_{B_\sigma}^J
                   - (I\leftrightarrow J) 
              \Big) 
\nonumber \\  
&=& x^I p^J - \frac{i}{2} \sum_{m \ne 0} \frac{1}{m}
    \Big( \alpha^I_{-m} \alpha^J_m
        + \tilde{\alpha}^I_{-m} \tilde{\alpha}^J_m \Big) 
    +\phi^I p_\phi^J - (I \leftrightarrow J).
\end{eqnarray}
\end{subequations}

\vspace*{-5mm} 
\noindent
Using the independent canonical variables (\ref{independent_variables}),  
the Poincar\'e algebra ISO($D-2, 2$) is satisfied, 
\begin{eqnarray}
\{P^I, P^J\} 
&=& 0, 
\nonumber \\
\{M^{IJ}, P^K\} 
&=& \eta^{IK} P^J - \eta^{JK} P^I, 
\label{poincare} \\
\{M^{IJ}, M^{KL}\} 
&=& \eta^{IK} M^{JL} - \eta^{JK} M^{IL}
  - \eta^{IL} M^{JK} + \eta^{JL} M^{IK}  , 
\nonumber
\end{eqnarray}
if the level matching condition $L_0^{\rm tr}=\tilde{L}_0^{\rm tr}$ 
is imposed. 
Conversely, the gauge fixing procedure we considered is the way 
to preserve the full $D$-dimensional Poincar\'e symmetry. 

According to the ordinary string theories in the light-cone gauge, 
we have to examine Poincar\'e algebra (\ref{poincare}) 
in the quantum theory~\cite{ggrt}. 
The checking of the Poincar\'e algebra is again straightforward, 
except for commutation relations 
$[M^{i-}, M^{j-}]$, $[M^{i\hat{-}}, M^{j\hat{-}}]$, 
and $[M^{i-}, M^{j\hat{-}}]$.  
After lengthy computation, 
we can obtain the following results,  
\begin{eqnarray}
{[}M^{i-}, M^{j-}{]} 
&=& \frac{4\pi{\phi^{\hat{+}}}^2}{(\phi^{\hat{+}}p^+-\phi^+p^{\hat{+}})^2} 
    A^{ij}, 
\nonumber \\
{[}M^{i\hat{-}}, M^{j\hat{-}}{]} 
&=& \frac{4\pi{\phi^+}^2}{(\phi^{\hat{+}}p^+-\phi^+p^{\hat{+}})^2}
    A^{ij}, 
\\
{[}M^{i-}, M^{j\hat{-}}{]} 
&=& i\delta^{ij}M^{-\hat{-}} 
    -\frac{4\pi\phi^{\hat{+}}\phi^+}{(\phi^{\hat{+}}p^+-\phi^+p^{\hat{+}})^2} 
    A^{ij}. 
\nonumber
\end{eqnarray}
An anomalous term $A^{ij}$ is 
\begin{eqnarray*}
A^{ij} 
&=& -2 \Big( 1-\frac{D-4}{24} \Big) 
       \sum_{m=1}^\infty m 
       \Big(  \alpha^i_{-m}\alpha^j_{m}
            + \tilde{\alpha}^i_{-m}\tilde{\alpha}^j_m
            \,- \,(i \leftrightarrow j)
       \Big)
\\
&&   + \,\Big( a_0 - \frac{D-4}{12} \Big)
       \sum_{m=1}^\infty \frac{1}{m}
       \Big(  \alpha^i_{-m} \alpha^j_m 
            + \tilde{\alpha}^i_{-m} \tilde{\alpha}^j_m
            \,- \,(i \leftrightarrow j)
       \Big) ,
\end{eqnarray*}
where the constant $a_0$ denotes the ordering ambiguity of the sum 
$L_0^{\rm tr}+\tilde{L}_0^{\rm tr}$ in (\ref{dependent_variables}) 
by adopting the normal-ordering prescription.  
The anomalous term $A^{ij}$ vanishes if and only if 
\begin{equation}
 D=28, \quad a_0=2.  
\label{critical_dimensions}
\end{equation}
Then, the Poincar\'e algebra ISO(26, 2) is satisfied 
in the quantum theory. 

A mass-shell relation of this string model is given by 
\begin{eqnarray}
m^2 
&=& -P^IP_I \nonumber \\
&=& 4\pi\Big(N+\tilde{N}-a_0\Big), 
\end{eqnarray}
where level operators $N$ and $\tilde{N}$ are defined by 
$$
N\equiv\sum_{m=1}^\infty\alpha^i_{-m}\alpha_{i m}, \qquad 
\tilde{N}\equiv\sum_{m=1}^\infty\tilde{\alpha}^i_{-m}\tilde{\alpha}_{i m}.
$$
The level matching condition $L^{\rm tr}_0 = \tilde{L}^{\rm tr}_0$ 
is then expressed as $N=\tilde{N}$. 
Therefore, this closed bosonic string model also 
involves tachyon in the ground state 
and graviton $g_{IJ}(\xi)$, two-form field $B_{IJ}(\xi)$ and 
dilaton $\phi(\xi)$ in the first excited massless state.    

%%%%%
\section{Conclusions and discussions}

\setcounter{equation}{0}
\setcounter{footnote}{0}
  
We have investigated the quantization of 
the {\UvUa} bosonic string model in two-dimensional quantum 
field theory.   
Even though the system has reducible and open gauge symmetries, 
we have shown that the covariant quantization has been successfully 
carried out in the Lagrangian formulation {\it \'a la} Batalin 
and Vilkovisky.  
In the covariant Batalin-Fradkin-Vilkovisky 
Hamiltonian formulation, 
we have considered the first-class constraints 
and the constraint algebra 
corresponding to the gauge symmetries and led to the same 
gauge-fixed action and BRST transformation  
as those of the Lagrangian formulation under the proper choice of 
the gauge-fermion and the identification of the fields. 
In addition we have obtained the BRST charge which generates 
the BRST transformations. 
Furthermore we have presented the noncovariant light-cone gauge 
formulation and investigated the symmetry of the background spacetime. 
With careful considerations of residual gauge symmetries,  
we have specified the gauge fixing conditions corresponding to the 
first-class constraints.   
Under these suitable conditions, 
we have been able to clarify dynamical independent variables 
and solve the first-class constraints explicitly. 
Although manifest covariance has been lost, 
we have confirmed the full $D$-dimensional Poincar\'e algebra of 
the background spacetime by direct computation.  

Since the quantizations of the model have been successfully 
carried out, we can argue the critical dimension of the string model. 
In our case, it turns to be 26+2. 
This means the background spacetime involves two time-like coordinates. 
Conversely, the requirement of two negative signatures in 
the background metric is natural one due to the gauge invariance 
of our model. 
The critical dimension has been obtained from both 
the BRST Ward identity in the BRST formulation 
and the $D$-dimensional quantum Poincar\'e algebra 
in the noncovariant light-cone gauge formulation. 
Therefore, we have concluded a consistent quantum theory of 
our {\UvUa} string model has only been formulated in 26+2-dimensional  
background spacetime. 
We have also considered the quantum states from the mass-shell relation. 
Contributions toward the mass-shell relation 
from zero-modes of the scalar field $\phi^I(x)$ are completely canceled, 
so that our closed bosonic string model possesses 
the same spectra as usual string theories. 

We propose the quantum {\UvUa} string model as a device to formulate 
the physics involving two time coordinates. 
In the formulation, the generalized Chern-Simons action 
has played an important role.   
From this viewpoint, it would be interesting to consider a low energy 
effective action which might be derived from our formulation of 
string theory.  
If we consider a background gauge field $A_I(\xi)$ 
which could be obtained from our open string or superstring model,   
it should have an additional gauge symmetry 
$\delta A_I(\xi)=\phi_I \Omega(\xi)$, 
where $\Omega(\xi)$ is a gauge parameter and $\phi_I$ is 
a constant null field, 
corresponding to the constraints (\ref{additional_constraints}).  
Such a gauge symmetry has been discussed in the formulation of 
10+2-dimensional supersymmetric Yang-Mills theory~\cite{ns}.  
In this context, 
the generalized Chern-Simons action~\cite{kawata}  
which can be formulated in arbitrary dimensions 
is also supposed to be useful 
for constructing the low energy effective action. 

The form of the classical action (\ref{classical_action_02}) 
suggests 
that this model should be more naturally defined in higher-dimensional 
field theories, namely, 
that membranes or $p$-branes are more fundamental 
than strings in our formulation. 
Actually, the action (\ref{classical_action_02}) is derived from 
a membrane action by adopting a compactification prescription. 
The {\UvUa} string model might be the first useful example which 
suggests higher-dimensional object 
like membrane or $p$-brane in the framework of perturbative 
field theory without using the concept of ``string duality''. 

\vspace{1cm}

%\newpage 

%%%%%
\noindent
{\Large{\bf Acknowledgments}}
\vspace{5mm}

We wish to thank J. Ambj{\o}rn, N. Kawamoto, H.B. Nielsen, 
K. Takenaga and H. Umetsu for interesting comments and discussions. 
We would like to thank Niels Bohr Institute for 
warm hospitality during the final stage of this work. 
We would also like to thank Dublin Institute for Advanced Studies 
for warm hospitality.    
The research of T.T. was partially supported 
by the Danish Rectors' Conference.  

\vspace*{1cm}

%%%%%
\noindent
{\large{\bf Appendix A. Two-dimensional world-sheet}}

%%%%%
\renewcommand{\theequation}{A.\arabic{equation}}
%%%%%

\setcounter{equation}{0}
\setcounter{footnote}{0}

\vspace*{5mm}

The two-dimensional spacetime coordinates are denoted by 
$x^m (=(\tau, \sigma))$. 
The two-dimensional flat metric $\eta_{mn}$ and 
Levi-Civit\'a symbol $\varepsilon_{mn}$ are given by  
\vspace*{1mm}
%
%%%%%
\renewcommand{\arraystretch}{1.0}
%%%%%
$$
\eta_{mn} =  
\eta^{mn} =  
\left( \begin{array}{cc}
         \ -1 \ & \ 0 \ \\
         \ 0  \ & \ 1 \
       \end{array}
\right),  
\qquad
\varepsilon_{mn} =  
- \varepsilon^{mn} =  
\left( \begin{array}{cc}
         \ 0 \ & \ 1 \ \\
         \ -1\ & \ 0 \
       \end{array}
\right). 
$$
%%%%%
\renewcommand{\arraystretch}{1.6}
%%%%%
%
\vspace*{3mm}
In the curved two-dimensional spacetime, 
the metric is given by $g_{mn}(x)$ 
and the covariant derivative $\nabla_m$ operates to fields as 
\begin{eqnarray*}
\nabla_m V_n &=& \del_m V_n - \Gamma^l{}_{mn} V_l, \\
\nabla_m V^n &=& \del_m V^n + \Gamma^n{}_{ml} V^l, 
\end{eqnarray*}
where $\Gamma^l{}_{mn}$ is the Christoffel symbol defined by 
$\Gamma^l{}_{mn} = 
 \frac{1}{2} g^{lk} (\del_m g_{kn} + \del_n g_{mk} - \del_k g_{mn})$.

The functional derivative with respect to 
a symmetric tensor $V^{mn}(x)=V^{nm}(x)$ is 
\begin{eqnarray*}
\frac{\delta V^{mn}(x)}{\delta V^{kl}(y)} = 
(\delta^m_k \delta^n_l + \delta^m_l \delta^n_k) \,\delta(x-y) .
\end{eqnarray*}
Then, the antibracket \rf{BRSTantibracket} is explicitly written as 
\begin{eqnarray*}
(X,Y) = 
\frac{\delta_{\rm R} X}{\delta \xi_I^*} 
\frac{\delta_{\rm L} Y}{\delta \xi^I} + 
\frac{\delta_{\rm R} X}{\delta \hat{A}_m^*} 
\frac{\delta_{\rm L} Y}{\delta \hat{A}^m} + \half 
\frac{\delta_{\rm R} X}{\delta \hat{g}_{mn}^*} 
\frac{\delta_{\rm L} Y}{\delta \hat{g}^{mn}} + \ldots .
\end{eqnarray*}

\vspace*{10mm}

\noindent
{\large{\bf Appendix B. Generalized Poisson bracket}}

%%%%%
\renewcommand{\theequation}{B.\arabic{equation}}
%%%%%

\setcounter{equation}{0}
\setcounter{footnote}{0}

\vspace*{5mm}

A generalized Poisson bracket~\cite{ht} is defined by 
$$
\{F, G\} \equiv
\bigg(\frac{\delta_{\rm L} F}{\delta\varphi^i}
      \frac{\delta_{\rm L} G}{\delta P_{\varphi^i}}
     -\frac{\delta_{\rm L} F}{\delta P_{\varphi^i}}
      \frac{\delta_{\rm L} G}{\delta\varphi^i}
\bigg) 
+(-)^{|F|}
\bigg(\frac{\delta_{\rm L} F}{\delta\theta^\alpha}
      \frac{\delta_{\rm L} G}{\delta P_{\theta^\alpha}}
     +\frac{\delta_{\rm L} F}{\delta P_{\theta^\alpha}}
      \frac{\delta_{\rm L} G}{\delta\theta^\alpha}
\bigg), 
$$
where canonical variables $\varphi^i$ and $P_{\varphi^i}$ are bosonic, 
and $\theta^\alpha$ and $P_{\theta^\alpha}$ are fermionic. 
In the above definition the contraction of the indices contains 
the integration of space or spacetime and $|F|$ is the Grassmann parity 
of $F$. 
This generalized Poisson bracket will be replaced 
by the graded commutation relation multiplied by $-i$ upon 
quantization, as usual. 
The explicit forms of the basic Poisson brackets are given by  
\begin{eqnarray*}
\{\varphi^i, P_{\varphi^j}\} 
&=& -\{P_{\varphi^j}, \varphi^i\} = \delta^i_j, \\
\{\theta^\alpha, P_{\theta^\beta}\} 
&=& \{P_{\theta^\beta}, \theta^\alpha\} = -\delta^\alpha_\beta. 
\end{eqnarray*}
The algebraic properties of the Poisson bracket are as follows: 
\begin{eqnarray*}
\{F, G\}
&=&-(-)^{|F||G|}\{G, F\}, \\
\{F_1F_2, G\} 
&=& F_1\{F_2, G\} + (-)^{|G||F_2|}\{F_1, G\}F_2.
\end{eqnarray*}

\vspace*{10mm}

\noindent
{\large{\bf Appendix C. BRST formulation of {\UvUa} model without
two-dimensional gravity}}

%%%%%
\renewcommand{\theequation}{C.\arabic{equation}}
%%%%%

\setcounter{equation}{0}
\setcounter{footnote}{0}

\vspace*{5mm}

In this appendix we summarize the Lagrangian 
BRST quantization of {\UvUa} model 
{\it without} two-dimensional gravity. 
Since the quantization of this model 
is much simpler than that of the model coupled with 
two-dimensional gravity we have investigated throughout this paper, 
the following result might be helpful to understand 
the quantization of the {\UvUa} gauge structure.   
   
The action of {\UvUa} model without two-dimensional gravity is 
\begin{eqnarray}
S = \int\! \dd^2 x \, 
    \bigg(&-& \frac{1}{2} \eta^{mn} \partial_m \xi^I \partial_n \xi_I 
          -  \eta^{mn} \partial_m \bar{\phi}^I \partial_n \phi_I \nonumber \\ 
         &+& \tilde{A}^m \phi_I \partial_m\xi^I 
          +  \tilde{B}^{mI} \partial_m \phi_I
          -  \frac{1}{2} \tilde{C} \phi^I \phi_I
    \,\bigg) ,
\label{classical_action_02_nogravity}
\end{eqnarray}
which is invariant under the following local gauge transformation, 
\begin{equation}
\begin{array}{rcl}
\delta\xi^I 
&=& v'\phi^I, 
\\
\delta \phi^I 
&=& 0, 
\\
\delta \bar{\phi}^I 
&=& u'^I, 
\\
\delta\tilde{A}^m 
&=& \ep^{mn} \del_n v + \eta^{mn} \del_n v', 
\\
\delta\tilde{B}^{mI} 
&=& \ep^{mn} \del_n u^I + \eta^{mn} \del_n u'^I
- v \ep^{mn} \del_n \xi^I + v' \eta^{mn} \del_n \xi^I 
- \tilde{w}^m \phi^I , 
\\
\delta\tilde{C} 
&=& \partial_m \tilde{w}^m 
+ \del_m v' \tilde{A}^m - v' \partial_m \tilde{A}^m .
\end{array}
\label{u1u1_nogravity} 
\end{equation}
After performing the BRST formulation,  
one obtains the following gauge-fixed action 
\begin{eqnarray}
S_{\rm gauge\hbox{-}fixed} &=& \int\! \dd^2 x \, \bigg\{
- \, \half \eta^{mn} \prt_m \xi^I \prt_n \xi_I 
- \eta^{mn} \prt_m \bar{\phi}^I \prt_n \phi_I 
- \eta^{mn} \prt_m \bar{f} \prt_n f
\nonumber\\&&\spaceSgf
- \hat{a}^m   ( \prt_m a   + \ep_m{}^n \prt_n a' )
- \hat{b}^m_I ( \prt_m b^I + \ep_m{}^n \prt_n b'^I )
\nonumber\\&&\spaceSgf
- \hat{c}^m   ( \prt_m c   + \ep_m{}^n \prt_n c' )
\nonumber\\&&\spaceSgf
- 2 a \tinyspace \hat{b}^m_I \prt_m \xi^I
+ \ep_{mn} \hat{b}^m_I \hat{c}^n \phi^I 
+ \half (f + a a') \ep_{mn} \hat{b}^m_I \hat{b}^{nI}
\, \bigg\}. 
\label{final_covariant_action_nogravity}
\end{eqnarray}
The action \rf{final_covariant_action_nogravity} is invariant under 
the nilpotent BRST transformations 
\begin{eqnarray}
s \xi^I &=&  a' \phi^I , 
\nonumber \\
s \phi^I &=& 0 , 
\nonumber \\
s \bar{\phi}^I &=&  b'^I - a' \xi^I , 
\nonumber \\
s f &=& 0 , 
\nonumber \\
s \bar{f} &=&  c' ,
\nonumber \\
s a &=& 0 , 
\nonumber \\
s a' &=& 0 , 
\nonumber \\
s b^I &=& ( f - a a' ) \phi^I , 
\label{final_BRSTPhi_nogravity} \\
s b'^I &=& 0 , 
\nonumber \\
s c &=&  \half \phi^I \phi_I,  
\nonumber \\
s c' &=& 0 , 
\nonumber \\
s \hat{a}^m &=& 
                  \ep^{mn} ( \phi_I \prt_n \xi^I - \prt_n \phi_I \xi^I ) 
                - a' \tinyspace \hat{b}^m_I \phi^I 
\nonumber \\
&&
                - \, ( \ep^{mn} c - \eta^{mn} c' ) \prt_n a'
                - \ep^{mn} \prt_n ( c a' + c' a ),
\nonumber \\
s \hat{b}^m_I &=& 
                    \ep^{mn} \prt_n \phi_I,
\nonumber \\
s \hat{c}^m &=& 
                  ( \ep^{mn} a - \eta^{mn} a' ) \prt_n a' 
                + \ep^{mn} \prt_n f .
\nonumber
\end{eqnarray}
%

%%%%%

\end{document}